\documentclass[useAMS,usenatbib]{mn2e}
\usepackage{graphicx}
\usepackage{subfigure}
\usepackage{amsmath}

\title[Assessing molecular line diagnostics]{Assessing molecular line diagnostics of triggered star formation using synthetic observations.}
\author[T. J. Haworth et al.]{Thomas J. Haworth$^{1,}$\thanks{E-mail:
haworth@astro.ex.ac.uk}, Tim J. Harries$^1$, David M. Acreman$^1$ and David A. Rundle$^2$\\
$^1$School of Physics, University of Exeter, Stocker Road, Exeter EX4 4QL\\
$^2$Met Office, FitzRoy Road, Exeter, EX1 3PB}
\begin{document}

\date{Accepted ???. Received ???; in original form ???}

\pagerange{\pageref{firstpage}--\pageref{lastpage}} \pubyear{2012}

\maketitle

\label{firstpage}

\begin{abstract}
We investigate {observational signatures of triggered star formation in bright rimmed clouds (BRCs)} by using molecular line transfer calculations {based on radiation-hydrodynamic radiatively-driven-implosion models.}
We find that for BRCs the separation {in velocity} between the line profile peak of an optically thick and an optically thin line is determined by both the observer viewing angle and the density of the shell driving into the cloud{. In agreement with observations, we find that most BRC line profiles} are symmetric and that asymmetries can be either red or blue, in contrast to the blue-dominance expected for a collapsing cloud.
Asymmetries in the {line profiles} arise when {an} optically thick line is dominated by the shell and {an} optically thin line is dominated by the cloud {interior to the shell}. {The asymmetries are} red or blue depending on whether the shell is moving towards or away from the observer {respectively}. 
{Using the known motions of the molecular gas in our models} we rule out the {`envelope expansion with core collapse' mechanism} as the cause of the lack of blue-asymmetry in {our simulated observations}. We show that the absence of a strong photon dominated region (PDR) around a BRC may not rule out the presence of triggered star formation{: if the BRC line} profile has a strong blue component then the shell is expected to be driving towards the observer, suggesting that the cloud {is being viewed} from behind and the PDR is obstructed. This could explain why BRCs such as {SFO 80, 81 and 86} have a blue secondary peak and only a weak PDR inferred at $8\,\mu\rm{m}$.
{Finally we also test the use of $^{12}$CO, $^{13}$CO and C$^{18}$O as diagnostics of
cloud mass, temperature and column density.} We find that the inferred conditions are in reasonable agreement with those from the models. Calculating the cloud mass assuming spherical symmetry is shown to introduce {an} error of an order of magnitude {whereas} integrating the column density over a given region is found to introduce an error of up to a factor of two.

\end{abstract}


\section{Introduction}
Molecular line diagnostics are a widely used tool for investigating the conditions of astrophysical clouds and star formation \citep[e.g.][]{1980ApJ...240...84S,2002ApJ...577..798D, 2006A&A...450..625U, 2008ApJS..177..341N, 2010MNRAS.407..986R, 2012MNRAS.422..521B, 2013MNRAS.429L..10H}. Line profiles can yield information about the kinematic motions of the molecular gas \citep[e.g.][]{2004JKAS...37..257L, 2008MNRAS.388..898T, 2010MNRAS.408.2426R, 2010MNRAS.407.2434S,2011MNRAS.412.1755L} and ratios of line intensities can be used to infer the cloud properties such as optical depth, temperature, column density and mass \citep[e.g.][]{1983ApJ...264..517M}. 

Bright rimmed clouds (BRCs) are objects which are believed to be formed when shocks generated by nearby massive stars drive into surrounding pre-existing density structures, potentially triggering star formation in the radiatively driven implosion (RDI) scenario \citep[e.g.][and references therein]{1982ApJ...260..183S,1989ApJ...346..735B,1994A&A...289..559L,2003MNRAS.338..545K,2009MNRAS.393...21G, 2009ApJ...692..382M,2010MNRAS.403..714M,2011ApJ...736..142B,2012A&A...538A..31T,2012MNRAS.420..562H}. This is a highly kinematic process in which shock driving is expected to occur, {potentially leading to the collapse of the cloud.} As such, BRCs have been subjected to a large {number} of molecular line observations to try and identify RDI \citep[e.g.][]{1997A&A...324..249L,2002ApJ...577..798D,2004A&A...414.1017T,2006A&A...450..625U,2009A&A...497..789U,2009MNRAS.400.1726M}. 

The conditions in the neutral gas of BRCs are frequently calculated based on the ratio of $^{13}$CO to C$^{18}$O intensities following \cite{1983ApJ...264..517M}. However alternative combinations of lines can also be used {such as CS, HCO$^{+}$, HCN and other CO isotopologues}. {\cite{1997A&A...324..249L} used CO, CS  and \cite{2004A&A...414.1017T} used $^{12}$CO, $^{13}$CO} to calculate the conditions of BRCs in IC1848. They then compared the neutral cloud conditions with the ionized boundary layer (IBL) pressure to determine whether or not {the clouds are} being compressed. \cite{2004A&A...414.1017T} found that two out of the three clouds studied have possibly been induced to collapse by the effect of radiation from nearby stars. The single system studied by \cite{1997A&A...324..249L} was also found to be in this state, where the IBL pressure was greater than the cloud support pressure. \cite{2006A&A...450..625U} performed a similar pressure comparison on {four} BRCs from the SFO catalogue \citep{1991ApJS...77...59S, 1994ApJS...92..163S} using $^{12}$CO, $^{13}$CO and C$^{18}$O (J\,=1\,$\rightarrow$\,0) transitions, finding probable triggering in two of the clouds. 
Other examples are \cite{2009MNRAS.400.1726M} and \cite{2009A&A...497..789U} where CO observations and signatures of photoionization were used to refine the northern and southern hemisphere SFO catalogues respectively, retaining only those clouds in which triggered star formation seems likely. {\cite{2009MNRAS.400.1726M} and \cite{2009A&A...497..789U} both} found that BRCs hosting sites of probable star formation typically had warmer external layers of neutral gas, approximately $20-30$\,K, compared to the central cloud which is at about $10-20$\,K. \cite{2009A&A...497..789U} retained clouds for which a photon dominated region (PDR) was clearly visible, a feature which suggests that photoionization is taking place. These gas studies have been reasonably successful in identifying possible sites of triggering, however none {has} been conclusive. Additional evidence and tests of the accuracy of the diagnostic techniques are still required before strong conclusions can be drawn about the prevalence of RDI.

A number of features have been identified in molecular line profiles that are believed to be characteristic of specific kinematic processes. For a gas undergoing Maxwell-Boltzmann thermal motions the line profile is described by a Gaussian distribution due to thermal broadening. Deviations from this form can give insight into the bulk motions of the molecular gas. A signature that is commonly interpreted as representing infall comes from observations of optically thick lines such as $^{12}$CO. If the optically thick line is sufficiently self-absorbed there will be two peaks in the line profile. For an infalling cloud, the red line profile peak  is due to material moving away from the observer in the exterior regions of the cloud and the blue line profile peak is due to material moving towards the observer in the central regions of the cloud. Given that the central regions are at higher density and are therefore more likely to exceed the critical density for the molecular species, a blue-asymmetry (when the blue peak is stronger) is the expected signature of infall \citep[e.g.][]{2004JKAS...37..257L, 2008MNRAS.388..898T, 2010MNRAS.408.2426R, 2010MNRAS.407.2434S}. A second, optically thin, line such as C$^{18}$O is checked for a single peak to ensure that the two peaks from the optically thick line are from the same cloud rather than a superposition of two distinct objects {at different systematic velocities}.

Although this blue-asymmetry has been observed for protostars \citep[e.g.][]{1997ApJ...489..719M} and pre-stellar cores \citep[e.g.][]{2001ApJS..136..703L,2004JKAS...37..257L} it is generally not observed in BRCs. Rather there is usually no clear asymmetry and sometimes even a dominant red-asymmetry \citep{2004A&A...419..599T}. For example \cite{2002ApJ...577..798D} use the Five College Radio Astronomy Observatory (FCRAO) to perform a number of molecular line observations of BRCs from the SFO catalogue. They found that a strong blue-asymmetry feature was only observed in one out of seven of the BRCs that they studied. \cite{2002ApJ...577..798D} proposed that this might be due to the shock heating the cloud from the outside in, which could render the standard infall signatures unobservable. 

It is currently unclear what is responsible for the lack of infall signature in the self-absorption peaks of optically thick line spectra of BRCs. There are a number of proposed causes, for example rotation \citep[e.g.][]{2004MNRAS.352.1365R}, pulsation \citep[e.g.][]{2006ApJ...652.1366K}, turbulence in the core \citep[e.g.][]{2009ApJ...699L.108L,2012ApJ...750...64S}, shock heating \citep{2002ApJ...577..798D} or the envelope expansion with core collapse (EECC) model \citep[e.g.][]{2006ApJ...652.1366K,2010MNRAS.403.1919G,2011MNRAS.412.1755L,2011ApJ...741..113F}. \cite{2004A&A...419..599T} studied the red-asymmetric BRC SFO 11NE in IC1848 and also attempt to model its line profile by calculating a synthetic profile for a number of possible cloud configurations. They found that an EECC model gave good agreement, suggesting that SFO 11NE is in the expansion phase of RDI identified by \cite{1994A&A...289..559L}.  

Understanding the reliability of {molecular line} diagnostics and the reason behind the lack of blue-asymmetry is essential if a more comprehensive picture of the effect of feedback and triggered star formation is to be realised. {In \cite{2012MNRAS.426..203H} we tested other diagnostics of BRCs that use continuum and atomic line data.}
In this paper we extend this form of analysis to molecular lines: generating synthetic data cubes and performing standard diagnostics to test their accuracy and applicability, and to address sources of ambiguity when using them to infer whether or not triggered star formation is occurring.

\section{Numerical method}
\label{num_meth}
We use the grid-based radiation transport and hydrodynamics code \textsc{torus} to perform non-LTE molecular line transfer calculations \citep[e.g.][]{2000MNRAS.315..722H,2010MNRAS.407..986R,2010MNRAS.406.1460A,2012MNRAS.420..562H}.
The details of the molecular line transfer algorithm are given in \cite{2010MNRAS.407..986R}. We perform a non-LTE statistical equilibrium calculation to determine the level populations and use the result to calculate synthetic observations in the form of spectral datacubes.

We use an accelerated Monte Carlo method \citep{2000A&A...362..697H} to calculate the mean intensity in each cell. A cell-centric long-characteristic ray tracing scheme is used in which a number of randomly directed rays are traced from random locations in each cell. The frequencies are also randomly selected from a uniform distribution of width 4.3 turbulent line widths, centred on the rest frequency of a given molecular transition. The specific intensity at the end point of the ray in the cell is determined by integrating the equation of radiative transfer along the path traced by the ray to the edge of the grid. 
The boundary condition for most rays is the cosmic microwave background (CMB). However for the calculations in this paper there is a nearby O star, the effect of which has to be included. A ray is therefore forced from each cell towards the star, using the stellar effective temperature as the boundary condition and weighting that ray's contribution based on the assumed probability of it having intersected the star at random. This probability is simply the solid angle subtended by the star divided by $4\rm{\pi}$.

We include dust in these calculations and assume a {canonical value for the} dust to gas mass ratio of $1\times10^{-2}$ in all cells that are below a temperature of 1500\,K. For cells hotter than this we incorporate sublimation effects by setting the dust abundance to a negligible value. We assume spherical silicate dust grains that follow a standard interstellar medium size distribution \citep{1977ApJ...217..425M}. The optical constants are taken from \cite{1984ApJ...285...89D}.  {Given that BRCs are relatively young (the RHD models simulated 200\,kyr of evolution), the dust size distribution and chemistry are not expected to depart much from this canonical interstellar medium model. This is the same dust treatment used in \cite{2012MNRAS.426..203H}.}

Once a set of rays has been traced, the mean intensity $\bar{J}_{\nu}$ in each cell is calculated by averaging the specific intensity from the rays, weighted by the line profile function
\begin{equation}
	\phi_{\nu} = \frac{c}{v_{\rm{turb}} \nu_0 \sqrt{\pi}}\rm{e}^{-\Delta \textit{v}^2/\textit{v}^2_{\rm{turb}}}
\end{equation}
where $c$, $\nu_0$, $v_{\rm{turb}}$, and $\Delta v$ are the speed of light, rest frequency of the transition, turbulent velocity and velocity required to Doppler shift $\nu_0$ to $\nu$ respectively. Here $v_{\rm{turb}}$ is imposed as 0.2\,km\,s$^{-1}$, {similar to that used in \cite{2010MNRAS.407..986R} and featured in \cite{2008ApJ...686.1174O}}. {The radiation hydrodynamic models of \cite{2012MNRAS.420..562H} (which are the basis for the statistical equilibrium calculations in this paper and are described more in section \ref{RHDModels}) exhibited strong systematic bulk motions which dominate turbulence.} $\bar{J}_{\nu}$ in each cell comprises two components, a first that is fixed for the cell during one set of level population iterations which describes the contribution from space external to the cell and a second that varies with the level populations (which affect the source function, $S_{\nu}$) internal to the cell
\begin{equation}
	\bar{J}_{\nu} = {J}_{\nu}^{\rm{ext}} +  {J}_{\nu}^{\rm{int}} = \frac{\sum\limits_i I_{\nu}^i\rm{e}^{-\tau_i}\phi_{\nu}}{\sum\limits_i \phi_{\nu}} + \frac{\sum\limits_i S_{\nu} \left(1-\rm{e}^{-\tau_i}\right)\phi_{\nu} }{\sum\limits_i \phi_\nu}
\label{jbar}
\end{equation}
where $I_{\nu}^i$ and $\tau_i$ are the intensity and optical depth along the $i^{\rm{th}}$ ray.
Equation \ref{jbar} is solved iteratively in conjunction with the equations of statistical equilibrium, which determine the level populations and modify the source function within the cell
\begin{eqnarray}
	\hspace{40pt}
	n_{l}\left[\sum\limits_{k<l} A_{lk} + \sum\limits_{k\neq l}\left(B_{lk}J_{\nu} + C_{lk}\right) \right] = \nonumber \\ \sum\limits_{k>l}n_kA_{kl} + \sum\limits_{k\neq l}n_k\left(B_{kl}J_{\nu} + C_{kl}\right)
\label{stateq}
\end{eqnarray}
where $n_l$, $A_{lk}$, $B_{lk}$ and $C_{lk}$ are the relative fractional level population of level $l$, Einstein A (spontaneous absorption/emission) and B (stimulated absorption/emission) coefficients and the collisional rate coefficient for levels $l$ and $k$ at a given temperature. The coefficients are taken from the LAMDA database \citep{2005A&A...432..369S}. Initially the $\rm{J} = 0, 1$ relative fractional level populations are set to 0.5 and the other levels to $1\times10^{-10}$.

The ray tracing and level population calculations are performed iteratively. Convergence is checked by comparing level populations from the latest and previous iteration, being achieved where the maximum root mean square fractional difference {in all levels} is less than a user-specified value, here taken to be $1\times10^{-2}$. {This results in an average fractional difference of order $10^{-4}$--$10^{-5}$ in the J = 2, 1 levels which are those required for the transitions used in this paper.} Some repeat calculations were performed to check {that our} convergence {criterion was sufficient}.

A two-stage calculation is performed in which an initial set of iterations using rays with fixed position, frequency and direction is run until the level populations converge. This is followed by iterations using rays with random position, frequency and direction that double in number until the level populations again converge. The first stage of the calculation converges quickly, {but} poorly samples both the frequency range and the spatial extent of the grid. The second stage reduces the systematic and random errors associated with using fixed rays. This combination reduces the calculation time compared to using solely random rays. A typical calculation requires around 10 iterations using {a starting number of between 400--700} fixed rays per cell {(that do not double in number between iterations)} followed by 3--4 iterations using random rays {which double in number with each iteration.} We also make use of the convergence-acceleration scheme of \cite{NgAcceleration}, which estimates an updated set of relative fractional level populations by extrapolation based on the level populations from the previous 4 iterations. This convergence acceleration is employed every five iterations. 

In this work we investigate molecular line diagnostics and kinematic signatures of the neutral component of BRCs. {Due to computational expense it currently not possible to perform 3D radiation hydrodynamic models with chemical evolution \citep{2010MNRAS.404....2G}. We therefore use standard values for the molecular abundance relative to hydrogen and neglect PDR and low--temperature chemistry other than to adopt the following conditions.} An abundance drop-model is employed at low temperatures for CO and its isotopologues to accommodate freeze-out of molecules on to dust grains \citep{2004A&A...424..589J}. Under this scheme molecular species in cells at less than 30\,K and molecular hydrogen density greater than $3\times10^{4}$\,cm$^{-3}$ have their abundance reduced by a factor of 10. The molecular abundance is set to a negligible value where the neutral atomic hydrogen fraction is lower than {the conservative value of} 0.1 as the gas is ionized and molecules would be dissociated {\citep[photodissociation codes assume no ionized atomic hydrogen, e.g.][]{2012MNRAS.427.2100B, 2013MNRAS.429.3584H}}.

Data cubes comprising two-dimensional spatial data and a series of velocity channels are generated by ray tracing in a similar manner to the primary level-population solver, only the rays are directed towards a pixel array that represents the image plane at the observer position. The cubes are produced in units of spectral radiance, erg\,s$^{-1}$\,cm$^{-2}$\,Hz$^{-1}$\,sr$^{-1}$, usually simply referred to as the monochromatic specific intensity $I_{\nu}$ and converted into a brightness temperature $T_{\rm{B}}$ using {the Rayleigh-Jeans approximation} 
\begin{equation}
	T_{\rm{B}} = \frac{I_{\nu}c^2}{2\nu^2k_{\rm{B}}}
\end{equation}
where $c$, $\nu$ and $k_{\rm{B}}$ are the speed of light, frequency of observation and Boltzmann constant respectively. 
Extensive testing of the molecular line transfer calculations is included in \cite{2010MNRAS.407..986R}.

\begin{figure}
\hspace{-10pt}
\includegraphics[width=9.6cm]{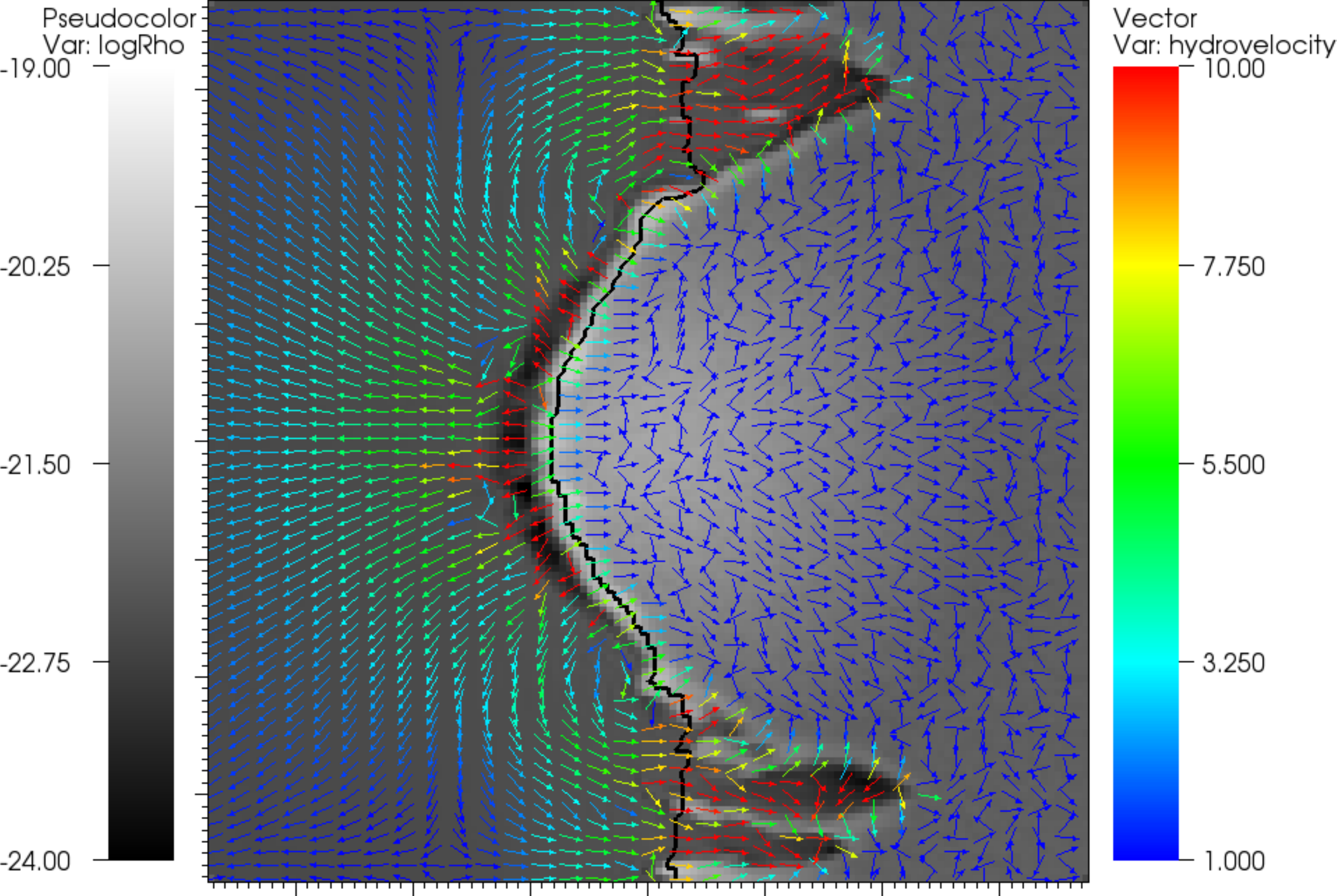}

\hspace{31pt}
\includegraphics[width=6.38cm]{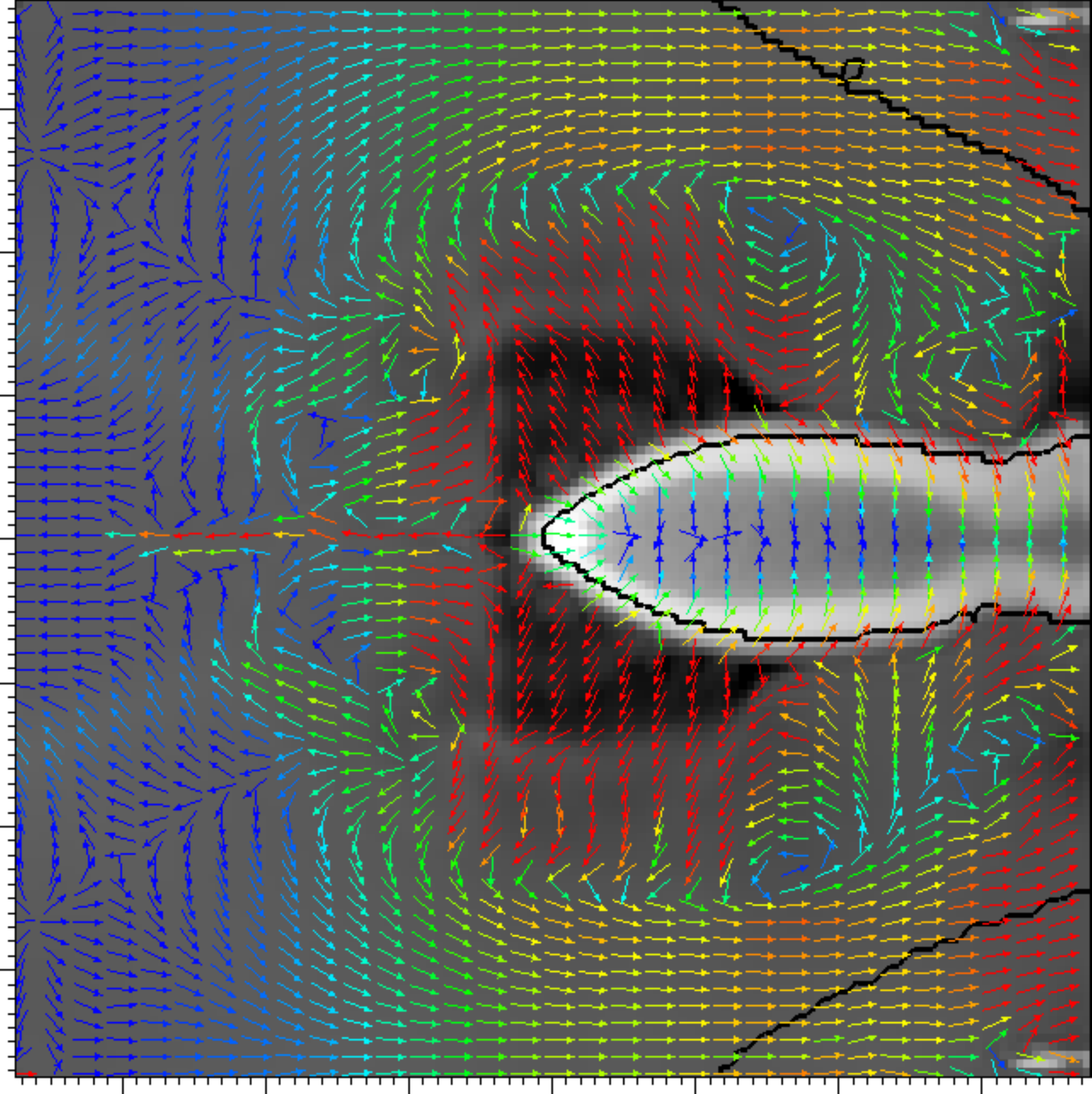}

\hspace{31pt}
\includegraphics[width=6.38cm]{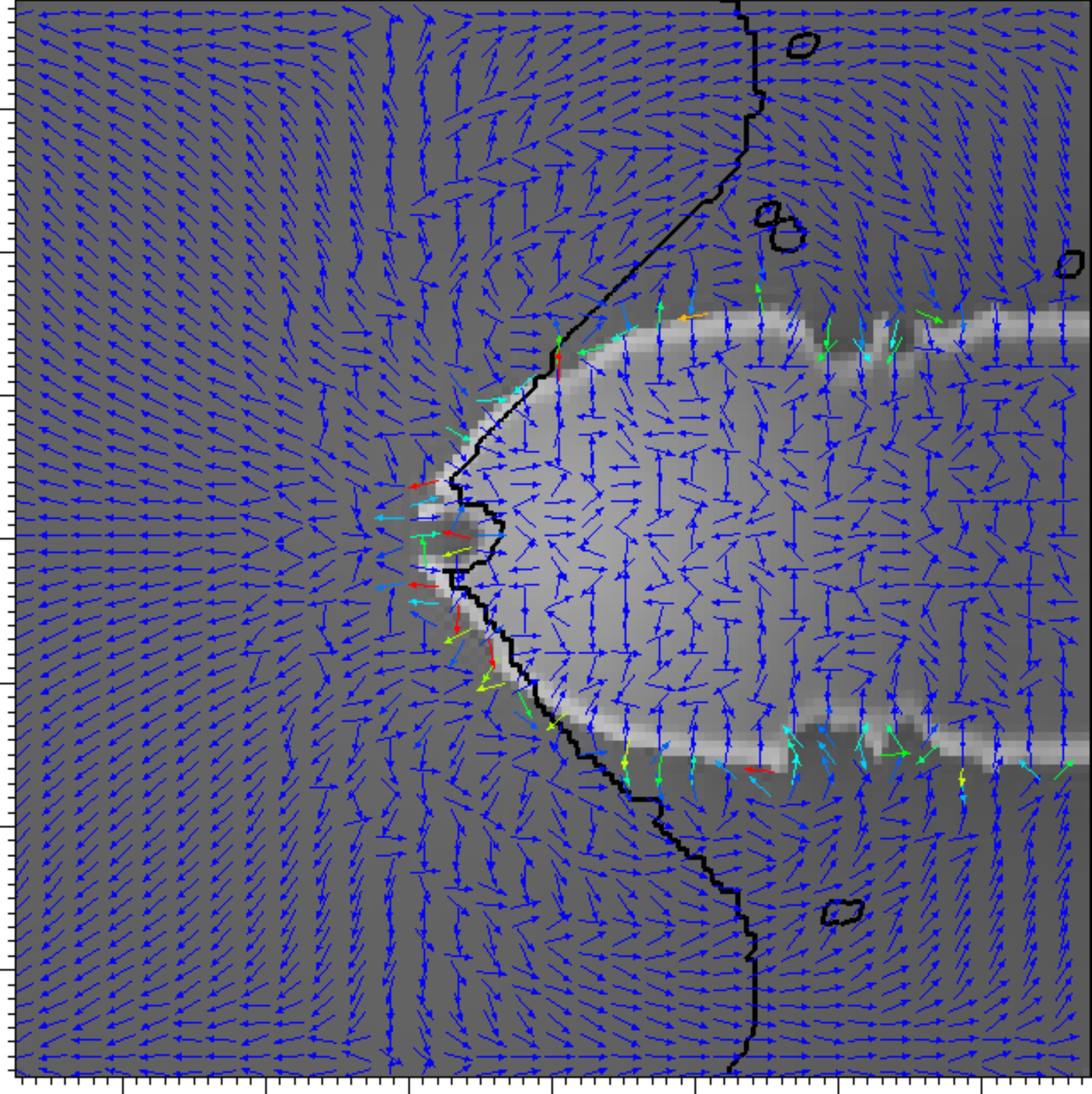}
\caption{{Slices through the final states of the RDI models from Haworth and Harries (2012). The panels are from the `low', `medium' and `high' flux models from top to bottom. The greyscale distribution is the logarithmic density, the contour is that at which the neutral atomic hydrogen fraction is equal to 0.1 and the vectors represent the velocity field. Major ticks are separated by 0.65\,pc.}}
\label{models}
\end{figure}

\subsection{The radiation hydrodynamic models}
\label{RHDModels}
The {density, temperature and velocity distribution that provide the basis for the statistical equilibrium and simulated observation calculations are taken from the final grid states of the radiation hydrodynamic RDI models of \cite{2012MNRAS.420..562H} that were also used in \cite{2012MNRAS.426..203H}.}

{In \cite{2012MNRAS.420..562H}, the models initially consisted of a Bonnor-Ebert sphere (BES) at the centre of the grid with a plane parallel ionizing radiation field impinging upon the left hand edge of the grid. We considered three different distances of the BES from the star responsible for the plane parallel radiation field which, due to the varying levels of flux incident at the left hand edge of the grid, we labelled the `low', `medium' and `high' flux models. The radiation field parameters are all given in Table \ref{RHparams}.  During the radiation hydrodynamic calculation, an ionization front was established which drove into the BES, accumulating a dense shell of material and changing the BES morphology. The outer layer of the shell was also photoionized and ejected in a photo-evaporative outflow. The manner in which compression of the BES proceeded was found to be dependent on the distance of the star, in agreement with previous models such as \cite{2009MNRAS.393...21G} and \cite{2011ApJ...736..142B}. We also found that inclusion of diffuse field radiation could significantly modify the result of the calculation. }

In this paper the final states of the most sophisticated {RDI} models, those which included the diffuse field {in \cite{2012MNRAS.420..562H} and have been subject to a full photoionization and thermal balance calculation in \cite{2012MNRAS.426..203H}}, are used. {A slice through the logarithmically scaled density distribution for each model at 200\,kyr (the simulation end time of the RDI models) is given in Figure \ref{models}. {Also included are} velocity vectors and a contour corresponding to the point at which the neutral atomic hydrogen fraction is equal to 0.1 (above which molecular species are able to survive, see section \ref{num_meth}). Where the contour does not trace the dense gas, for example in the wings of the high flux model (the bottom panel of Figure \ref{models}) the additional cooling from the full photoionization and thermal balance calculation in \cite{2012MNRAS.426..203H} has moved the ionization front. This relocation of the ionization front has a negligible effect on the simulated molecular line diagnostics in this paper since it is in regions away from the main cloud (the object of study) that are at low density and therefore low intensity relative to the cloud.}

\begin{table}
\caption{{A list of key parameters from the RDI models of  Haworth \& Harries (2012).}}
\label{RHparams}
\hspace{-25pt}
\begin{tabular}{c c l}
\hline
Variable (unit) & Value & Description  \\
\hline
 $R_{\rm{c}}$ (pc) & 1.6 & Cutoff radius of initial BES\\
 $n_{\rm{max}}$ ($\textrm{cm}^{-3} $)& $1000$ & Peak initial BES number density\\
 $\Phi_{\rm{low}}$ ($\textrm{cm}^{-2}$) & $9.0\times10^8$& Low ionizing flux\\
 $D_{\rm{low}}$ (pc) & $(-10.679,0,0)$ & Source position (low flux) \\
 $ \Phi_{\rm{med}}$ ($\textrm{cm}^{-2}$) & $4.5\times10^9$ & Intermediate ionizing flux\\
 $D_{\rm{med}}$ (pc) & $(-4.782,0,0)$ & Source position (medium flux) \\ 
 $\Phi_{\rm{high}}$ ($\textrm{cm}^{-2}$) & $9.0\times10^9$ & High ionizing flux\\ 
 $D_{\rm{high}}$ (pc) & $(-3.377,0,0)$ & Source position (high flux)\\
 $T$ (K) & 40000 & Source effective temperature\\
$R$ (R$_{\odot}$) & 10 & Source radius \\
 L(pc$^3$) & $4.87^3$ & Grid size\\
\hline
\end{tabular}
\end{table}

\section{Simulated Observations}
\label{sim_obs}
\subsection{Choice of molecular transitions}
\label{trans}
Isotopologues of CO are among the most commonly used species in molecular line observations, in particular for observations of BRCs \citep[e.g.][]{1997A&A...324..249L,2002ApJ...577..798D,2004A&A...414.1017T,2009MNRAS.400.1726M}. This is because they have a relatively high abundance and low critical density, making them easier to observe. We therefore choose to generate data cubes of $^{12}$CO, $^{13}$CO and C$^{18}$O. Analysis of molecular line data requires the use of probes which are sensitive to different conditions in the cloud. $^{12}$CO (J\,=\,2$\rightarrow$1) is a line which can be optically thick, with $^{13}$CO (J\,=\,2$\rightarrow$1) and C$^{18}$O (J\,=\,2$\rightarrow$1) being optically thinner variants. These are combined to trace and infer the properties of the molecular gas.

\subsection{Simulated instruments}
We smooth the data cubes that we calculate to a Gaussian beam using \textsc{aconvolve} from \textsc{ciao} v4.1 \citep{2006SPIE.6270E..60FB} to a size appropriate to the half power beamwidth (HPBW) of the simulated instrument.
We choose {the beam size} given by the Rayleigh criterion, as is the case for the JCMT which has a 15\,m dish, resulting in beam sizes of around $22\arcsec$. 
For comparison, the beam size of the  $^{12}$CO (J\,=\,1$\rightarrow$0) transition using the FCRAO in \cite{2002ApJ...577..798D} is $46\arcsec$. Factors such as instrument and atmospheric noise are not included. 


\section{Calculating the molecular cloud conditions}
\label{calcCond}
The optical depth, excitation temperature and column density of the optically thin C$^{18}$O line can be determined following the method described by \cite{1983ApJ...264..517M} and used by, for example, \cite{2004A&A...428..723U} and \cite{2009MNRAS.400.1726M}. The source-averaged optical depth of C$^{18}$O is determined using 
\begin{equation}
	\frac{T_{13}}{T_{18}} = \frac{1-{e}^{-\tau_{13}}}{1-{e}^{-\tau_{18}}}
	\label{tauFromT}
\end{equation}
where $T_{13}$ and $T_{18}$ are the peak brightness temperatures of the source-averaged line profiles of $^{13}$CO and C$^{18}$O respectively, with the background signal subtracted. In this work, the background signal is determined from the average value in the ambient H\,\textsc{ii} region. Equation \ref{tauFromT} is solved by assuming that the two source-averaged optical depths are related by their relative abundances
\begin{equation}
	\tau_{13}=X_{13/18}\tau_{18}
	\label{oDepths}
\end{equation}
where $X_{13/18}$ is the ratio of $^{13}$CO to C$^{18}$O abundances. This abundance ratio is usually estimated based on {Galactic} abundance distributions {and the location of the target in the Galaxy} \citep{1990ApJ...357..477L}, for example being taken as 10 in \cite{2006A&A...450..625U}. Here we use the ratio of prescribed abundances from Table \ref{params}, giving a ratio of approximately 16 for $X_{13/18}$ and 30 for $X_{12/13}$. We find the remaining single unknown optical depth numerically, using a decimal search.

The gas excitation temperature is estimated for $^{12}$CO and C$^{18}$O via the same approach used in \cite{2009MNRAS.400.1726M}. The equation of radiative transfer written in terms of optical depth $\tau$, integrated along a path length $s$ and with the background subtracted gives the intensity as a function of the Planck function $B_{\nu}$, the background intensity $I_0$ and the optical depth
\begin{equation}
	I_{\nu}(s) = \left( B_{\nu} - I_0 \right)\left(1-{e}^{-\tau}\right).
	\label{radtrans}
\end{equation}
This can be re-written in terms of temperatures as
\begin{equation}
T_{\rm{B}} = \frac{h\nu}{k_{\rm{B}}}\left(\frac{1}{{e}^{h\nu/k_{\rm{B}}T_{\rm{ex}}} - 1} - \frac{1}{{e}^{h\nu/k_{\rm{B}}T_{\rm{cmb}}}} \right)\left(1-{e}^{-\tau} \right)
	\label{T_B}
\end{equation}
where $T_{\rm{B}}$, $T_{\rm{ex}}$ and $T_{\rm{cmb}}$ are the brightness, excitation and CMB temperatures respectively and $\nu$ is the frequency of radiation emitted following the molecular transition \citep{1996tra..book.....R}. Equation \ref{T_B} can be rearranged for the excitation temperature to
\begin{equation}
T_{\rm{ex}} = T_{\nu}\left\{\ln\left[1+ T_{\nu}\frac{1-{e}^{-\tau}}{T_{\rm{B}} + T_{\nu}\left(1-{e}^{-\tau}\right) {e}^{\frac{-T_{\nu}}{T_{\rm{cmb}}}}}\right] \right\}^{-1}
	\label{Texcit}
\end{equation}
where $T_{\nu}$ is set to $h\nu/k_{\rm{B}}$.
If the $^{12}$CO emission is optically thick, which is typically expected to be the case, the term $1-e^{-\tau}$ tends to $1$ and the excitation temperature $T_{\rm{ex}}$ of $^{12}$CO can be estimated using
\begin{equation}
	T_{\rm{ex}} = 11.06\left\{\ln\left[1+11.06\frac{1}{T_{\rm{B}} + 0.192}\right] \right\}^{-1}.
\label{12coExciT}
\end{equation}
The criterion that $^{12}$CO be optically thick can be checked by comparing the relative intensities of $^{12}$CO and $^{13}$CO to their relative abundances. If the {$^{12}$CO to $^{13}$CO line intensity ratio is much smaller than the assumed abundance ratio then $^{12}$CO is expected to be optically thick.}

Assuming {local thermodynamic equilibrium} (LTE) and that a single temperature $T$ applies to the whole cloud, the kinematic temperature is simply this excitation temperature. \cite{2009MNRAS.400.1726M} derived the C$^{18}$O excitation temperature of clouds in addition to $^{12}$CO and found significant differences. Since $^{12}$CO and C$^{18}$O probe different parts of the cloud it is not surprising that they will have different excitation temperatures. Typically the interior parts of the cloud, probed by C$^{18}$O, are expected to be cooler. With its optical depth known, the C$^{18}$O excitation temperature can be calculated independently using equation \ref{Texcit}, giving
\begin{equation}
	T_{\rm{ex}} = 10.54\left\{\ln\left[1+\frac{10.54\left(1-{e}^{-\tau_{18}}\right)}{T_{\rm{B}} + 0.221\left(1-{e}^{-\tau_{18}}\right)}\right] \right\}^{-1}.
\label{c18oExciT}
\end{equation}
The total column density for a given molecular species is calculated following the method given in \cite{1986ApJ...303..416S}, whereby the optical depth is integrated over the line profile. For CO molecules, assuming a rigid rotor and that all levels are represented by a single excitation temperature, the column density over all levels is
\begin{equation}
	N = \frac{3k_{\rm{B}}}{8\pi^3B\mu^2}\frac{{e}^{\textit{hB}\rm{J}_\textit{l}(\rm{J}_\textit{l}+1)/\textit{k}_{\rm{B}}\textit{T}_{\rm{ex}}}}{(\rm{J}_\textit{l}+1)}\frac{T_{\rm{ex}}+hB/3k}{\left(1-{e}^{-\textit{h}\nu/\textit{k}_{\rm{B}}\textit{T}_{\rm{ex}}}\right)}\int \tau_{v} dv
	\label{Ntot}
\end{equation}
where $B$ and $\mu$ are the rotational constant and permanent dipole moment of the molecule respectively. $\rm{J}_\textit{l}$ is the lower of the two rotational levels for the transition being considered. The rotational constant and permanent dipole moment of C$^{18}$O are $54.891$\,GHz and 0.11 Debye respectively. When using equation \ref{Ntot} in this paper we use the average value of $\tau_{18}$ calculated via Equations \ref{tauFromT} and \ref{oDepths} and use the FWHM of the line to replace the velocity integral. 
With the C$^{18}$O column density known, the H$_2$ column density $N\left(\rm{H}_2 \right)$ can then be found using an assumed (and, since in this case it is prescribed, correct) abundance of C$^{18}$O relative to H$_2$, namely $1.7\times10^{-7}$ {\citep[the prescribed value is given in Table \ref{params} and was taken from][]{1997IAUS..170..113G}}. 

Following \cite{2006A&A...450..625U} {(assuming spherical symmetry)} we estimate the cloud average number density $n_{\rm{H}_2}$ using the column density via
\begin{equation}
	n_{\rm{H}_2} = \frac{\pi^3 N\left(\rm{H}_2 \right)}{8R}
	\label{numdens}
\end{equation}
where $R$ is the cloud radius. Equation \ref{numdens} is derived by integrating the column density over the assumed uniform density sphere and dividing by the circular surface presented to the observer.
Finally the mass of the cloud can be estimated using 
\begin{equation}
         M_{\rm{cloud}} = \frac{4\pi R^3}{3}n_{\rm{H}_2}\mu m_{\rm{H}}
	\label{M_sph}
\end{equation}
where $R$, $m_{\rm{H}}$ and $\mu$ are the cloud radius, atomic hydrogen mass and mean molecular weight respectively. We follow \cite{2006A&A...450..625U} and use a value of $\mu = 2.3$, which assumes 25 per cent abundance of helium by mass. 

In addition to the above mass calculation which assumes spherical symmetry and uniform density, the mass can also be calculated by integrating the inferred column density over a given solid angle. That is, the total mass of {molecular hydrogen} of over all pixels $i$ in a given region is
\begin{equation}
	M_{\rm{cloud}} = \frac{\mu m_{\rm{H}}}{X(j)}\sum_i a_i N_i(j)
	\label{M_int}
\end{equation}
where $X(j)$ is the abundance of species $j$ relative to molecular hydrogen and $a_i$ is the pixel area.

\section{Results and discussion}

\subsection{Synthetic data cubes}
\label{datCube}

\subsubsection{Cube construction}
We generated $^{12}$CO, $^{13}$CO and C$^{18}$O data cubes from the results of statistical equilibrium calculations in the manner described in sections \ref{num_meth} and \ref{sim_obs}. A total of nine statistical equilibrium calculations were run, one for each isotopologue considered at the three distances of the ionizing star from the cloud {(see section \ref{RHDModels} for details of the model)}. The maximum velocity magnitude in the data cubes is 10\,km\,s$^{-1}$ and each velocity channel spans 0.1\,km\,s$^{-1}$.  A summary of the statistical equilibrium and datacube calculation parameters is given in Table \ref{params}. The abundance of each species in Table \ref{params} is a constant value relative to H$_2$, determined through a literature search.

\subsubsection{Edge-on morphology}
\begin{figure}
\includegraphics[width=9.25cm]{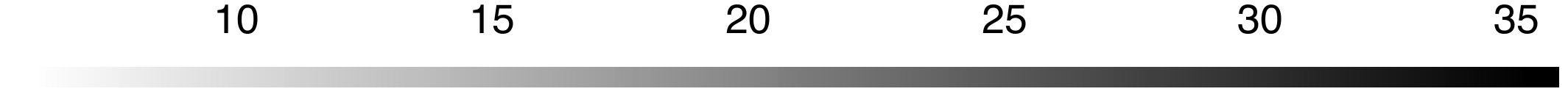}
\includegraphics[width=9.25cm]{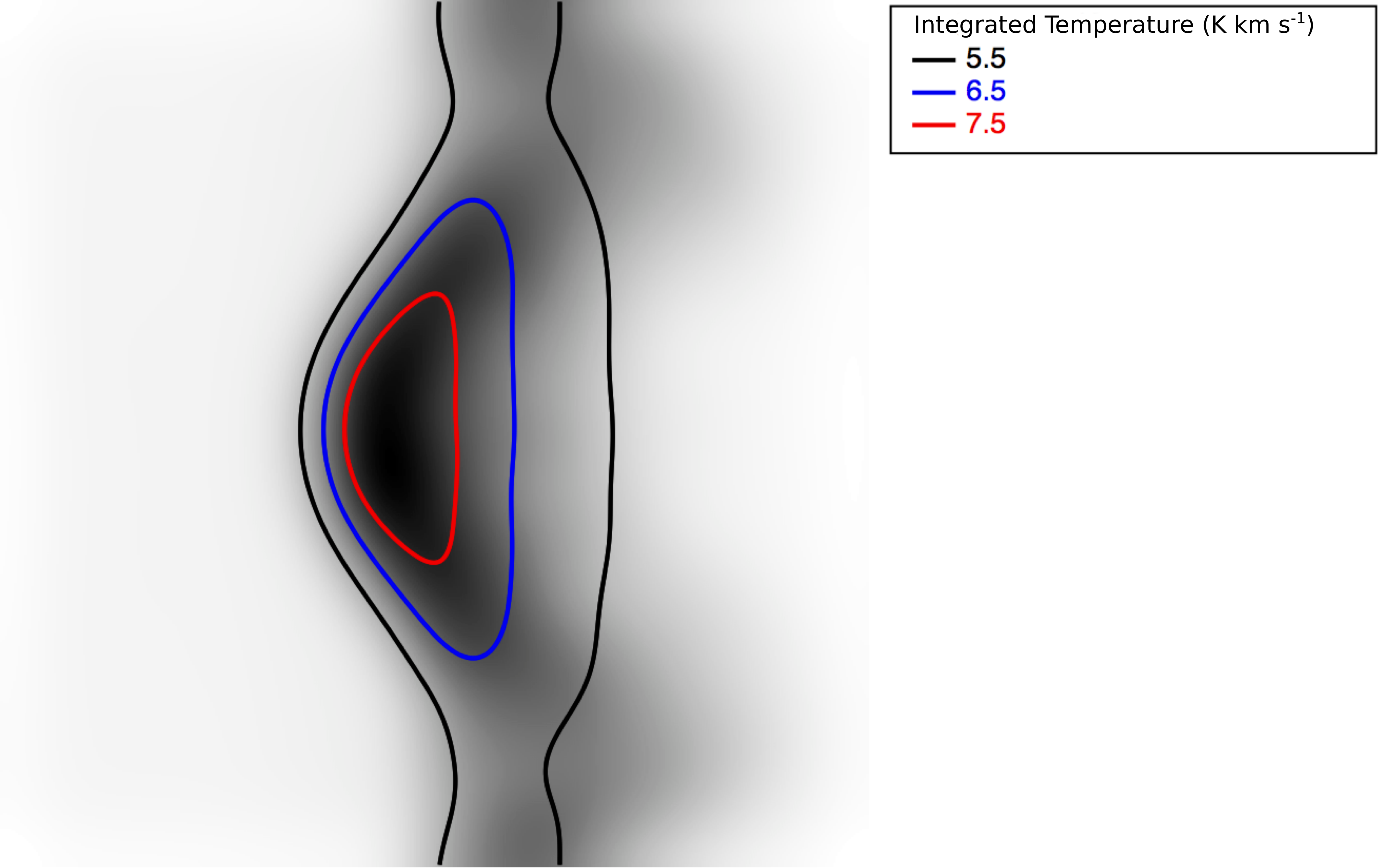}
\includegraphics[width=9.25cm]{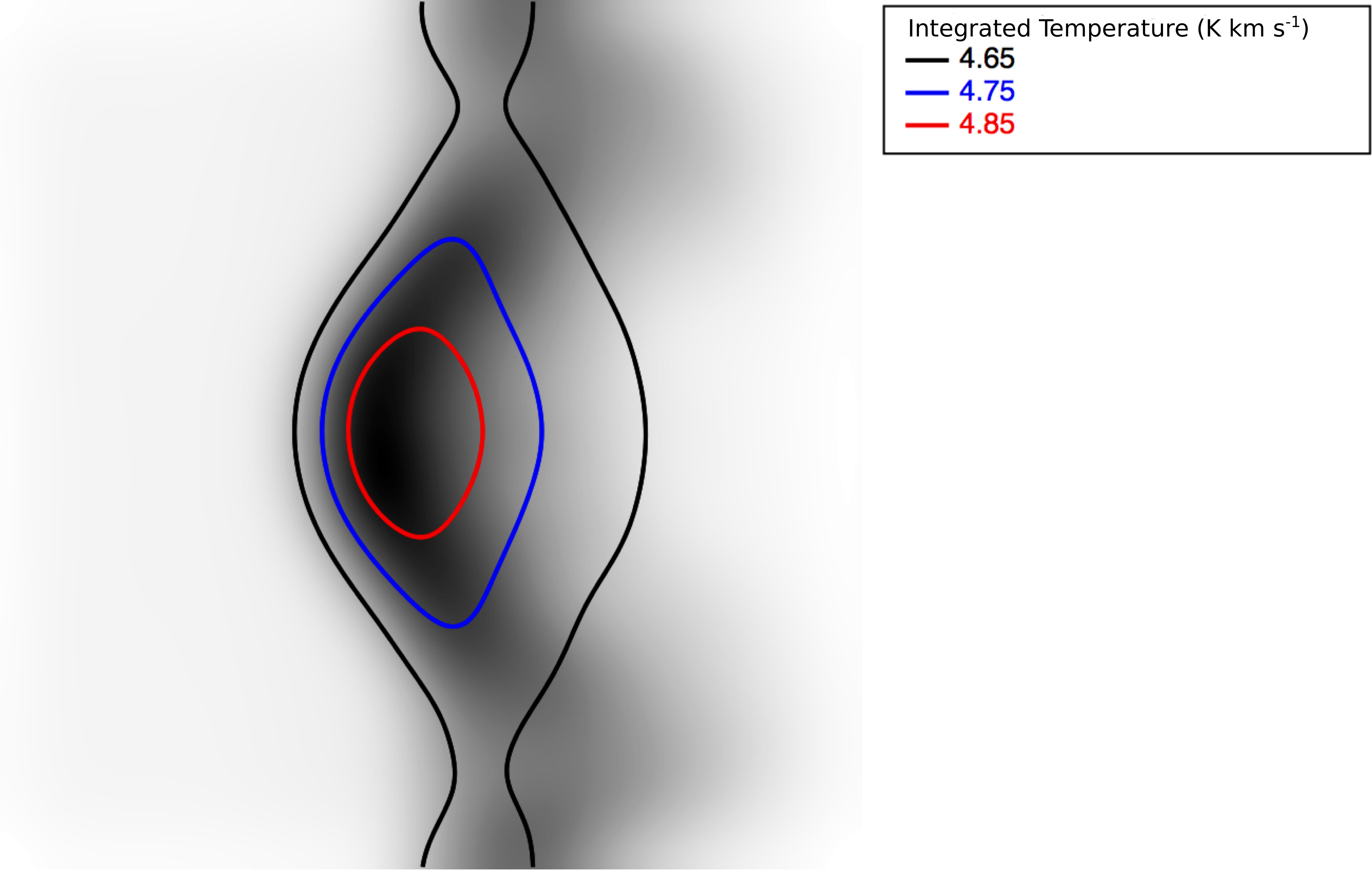}
\caption{Low flux model data cube $^{12}$CO images (greyscale) smoothed to a Gaussian beam representative of the JCMT {and integrated over velocity channels}. The image temperature scale is given by the greyscale bar in integrated brightness temperature units {(K\,km\,s$^{-1}$)}. The contours are $^{13}$CO (top) and C$^{18}$O (bottom). These images are all 4.87\,pc to a side.}
\label{molImgs_low}
\end{figure}

\begin{figure}
\hspace{-7pt}
\includegraphics[width=9.25cm]{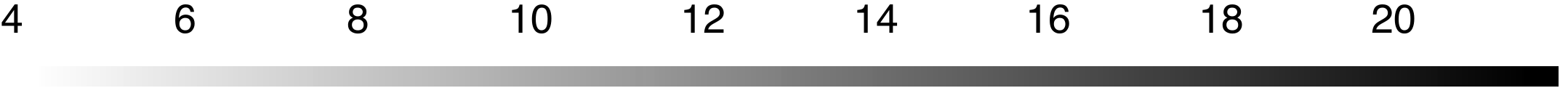}

\hspace{-7pt}
\includegraphics[width=9.25cm]{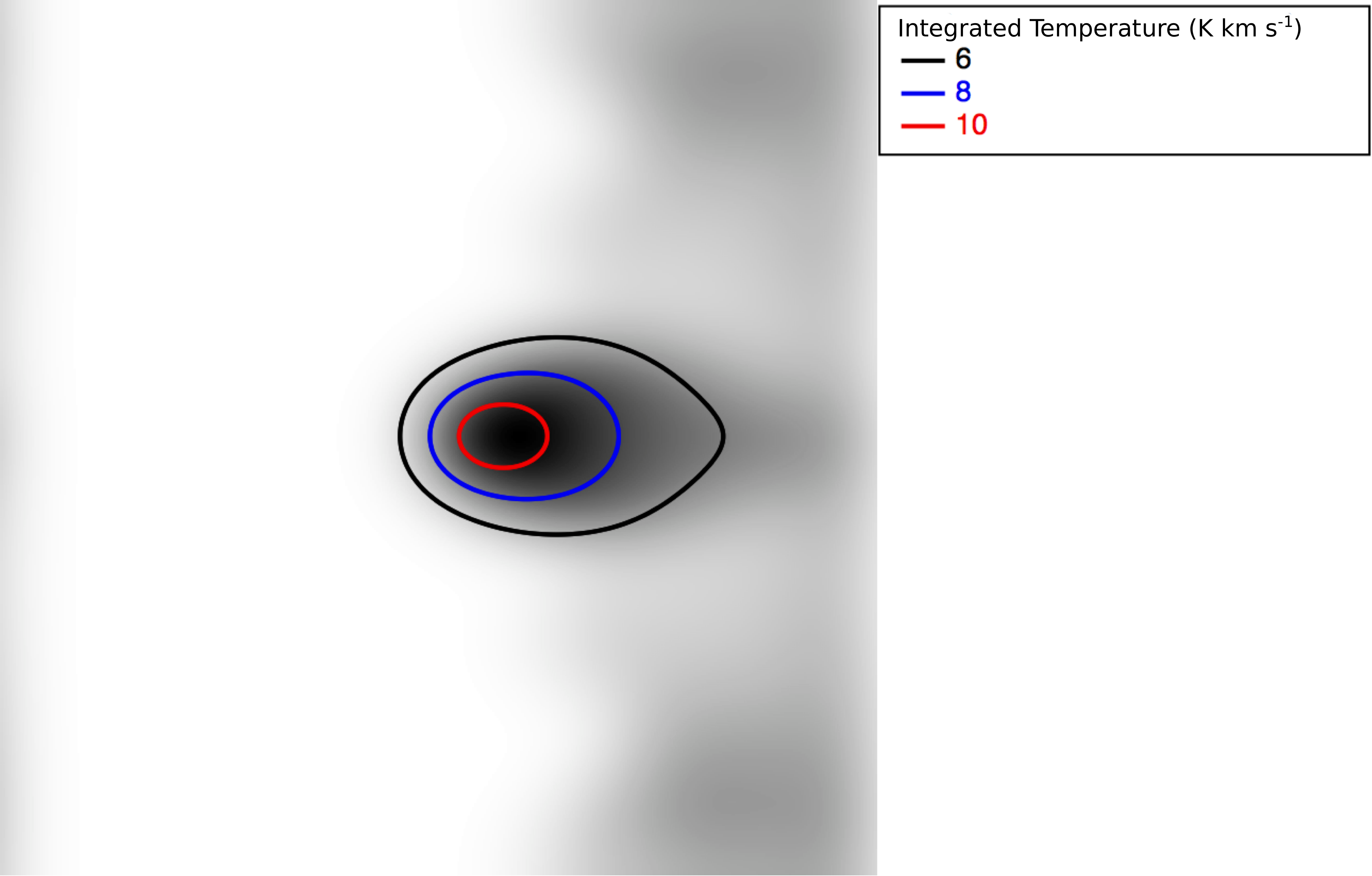}

\hspace{-7pt}
\includegraphics[width=9.25cm]{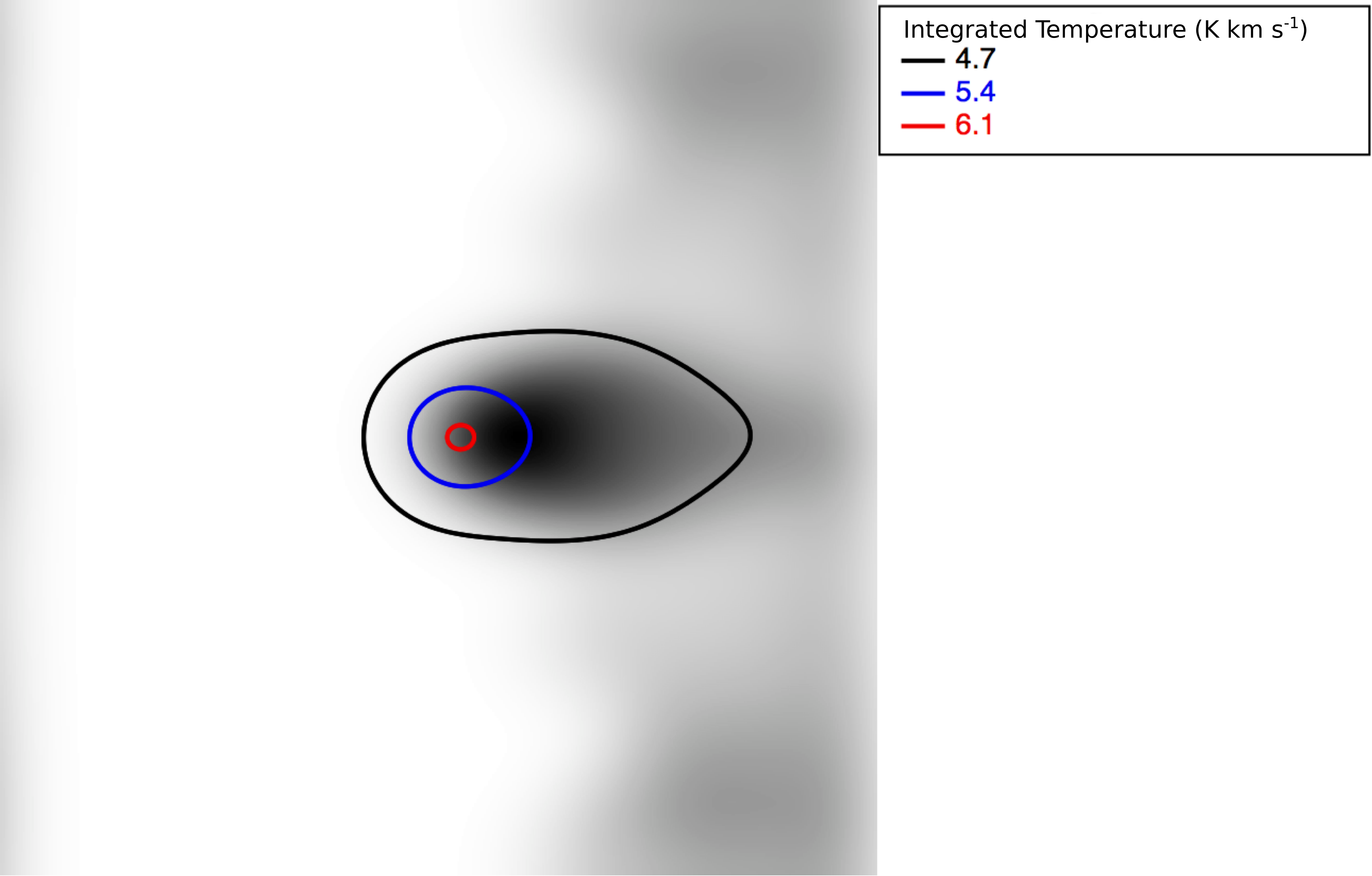}
\caption{Medium flux model data cube $^{12}$CO images (greyscale) smoothed to a Gaussian beam representative of the JCMT {and integrated over velocity channels}. The image temperature scale is given by the greyscale bar in integrated brightness temperature units {(K\,km\,s$^{-1}$)}. The contours are $^{13}$CO (top) and C$^{18}$O (bottom). These images are all 4.87\,pc to a side.}
\label{molImgs_med}
\end{figure}

\begin{figure}
\includegraphics[width=9.25cm]{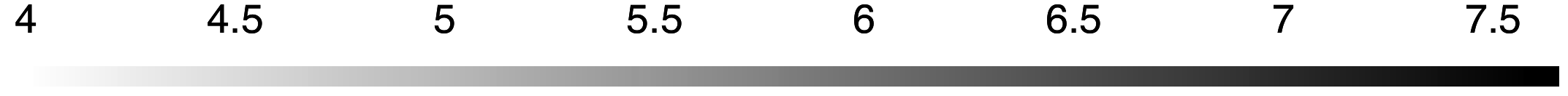}
\includegraphics[width=9.25cm]{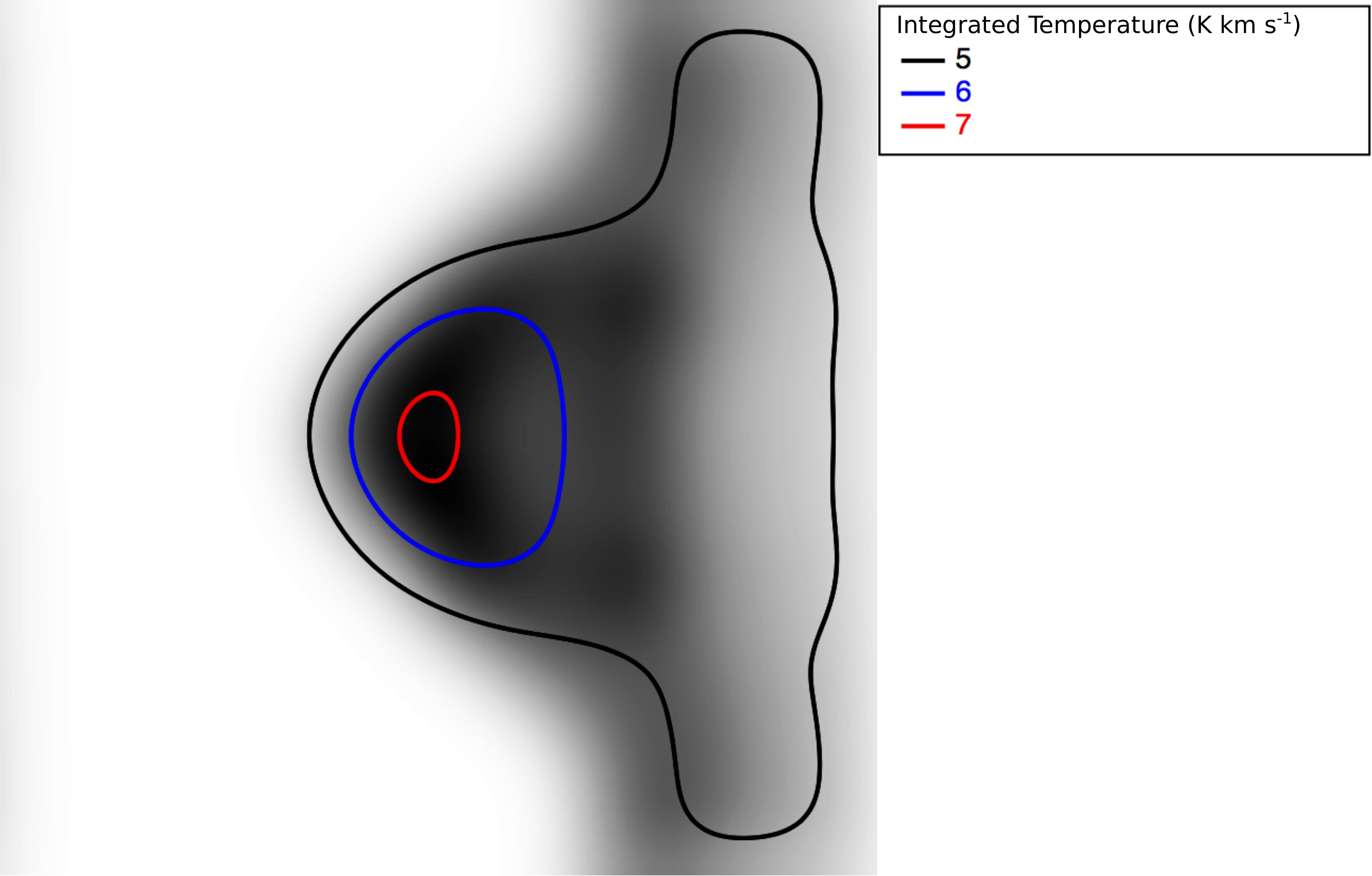}
\includegraphics[width=9.25cm]{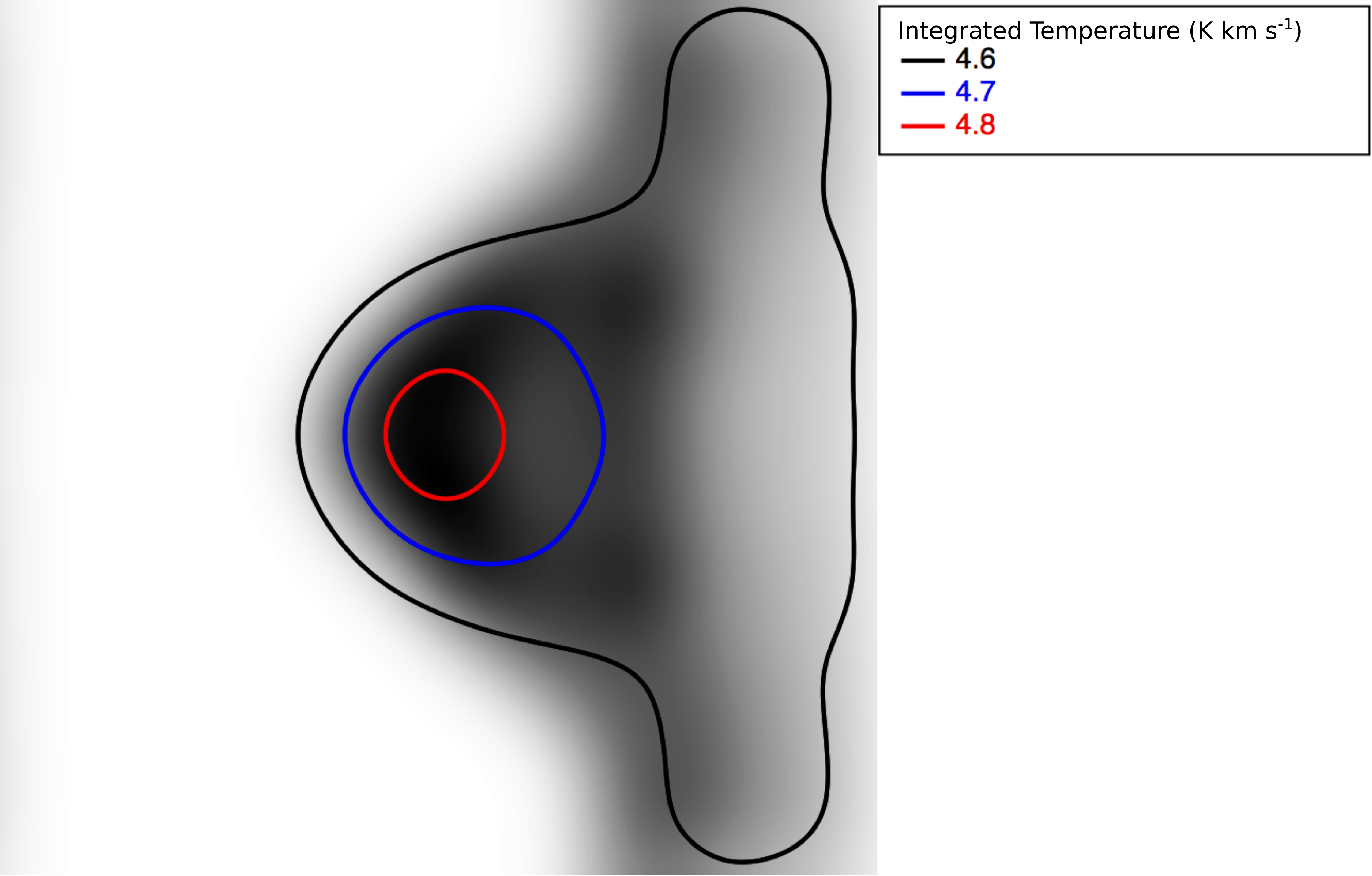}
\caption{High flux model data cube $^{12}$CO images (greyscale) smoothed to a Gaussian beam representative of the JCMT {and integrated over velocity channels}. The image temperature scale is given by the greyscale bar in integrated brightness temperature units {(K\,km\,s$^{-1}$)}. The contours are $^{13}$CO (top) and C$^{18}$O (bottom). These images are all 4.87\,pc to a side.}
\label{molImgs_hi}
\end{figure}

{$^{12}$CO} images of the clouds for an observer edge-on to the BRC, convolved to the appropriate Gaussian beam size, are given in Figures \ref{molImgs_low}, \ref{molImgs_med} and \ref{molImgs_hi} for the low, medium and high flux models respectively. Overlaid are $^{13}$CO and  C$^{18}$O {intensity} contours. These images and contours are constructed by integrating the data cubes using the \textsc{starlink} software \textsc{gaia}. The contours are chosen to give the best representation of the distribution of emission throughout the BRCs. The ionizing star is located off the left hand edge of the images.

\begin{table*}
\caption{The parameters used in statistical equilibrium and datacube generation calculations}
\label{params}
\begin{tabular}{c c l}
\hline
Parameter (Unit) & Value & Description \\
\hline
$v_{\rm{turb}}$ (km\,s$^{-1}$) & 0.2 & Turbulent velocity \\
$N_{\rm{v}}$ & 200 & Number of velocity channels \\
d$v$ (km\,s$^{-1}$) & 0.1 & Span of each velocity channel \\
$N_{\rm{pix}}$ & $401^2$ & Number of pixels per channel \\
$\theta$ ($\arcsec$) & 2.5 & Angular width per pixel \\
tolerance & $1\times10^{-2}$ & Statistical equilibrium convergence checking tolerance \\
$n_{\rm{C}^{18}\rm{O}}/n_{\rm{H}_2}$ & $1.7\times10^{-7} $& C$^{18}$O abundance \citep{1997IAUS..170..113G} \\
$n_{^{13}\rm{CO}}/n_{\rm{H}_2}$ & $2.7\times10^{-6}$& $^{13}$CO abundance \citep[][and references therein]{2008ApJ...679..481P} \\
$n_{\rm{^{12}CO}}/n_{\rm{H}_2}$ & $8.0\times10^{-5}$ & $^{12}$CO abundance \citep[][and references therein]{1988ApJ...326..909M}\\
$D$ (pc) & 1000 & Distance of observer \\
\hline
\end{tabular}
\end{table*}

The low flux $^{12}$CO images (Figure \ref{molImgs_low}) are dominated by the bright bow, with weaker emission in the wings of the cloud. There is also some weaker emission behind the bow in the cloud. The integrated $^{13}$CO and C$^{18}$O contours trace the $^{12}$CO morphology well. The peak C$^{18}$O contours extend further into the neutral gas away from the bright rim due to the lower {critical density} of the line. 
In the radiation hydrodynamic calculations instabilities arose resulting in fingers with dense tips in the wings of the BRC \citep{2012MNRAS.420..562H}. Those dense tips are not individually resolved in these data cubes due to the beam size used, rather they appear to contribute to the more widespread emission in the wings of the BRC.

\begin{figure*}
\includegraphics[width=17.5cm]{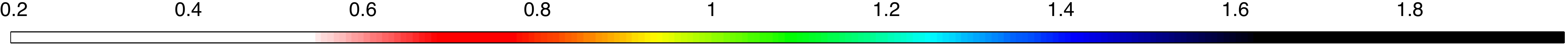}
\includegraphics[width=17.5cm]{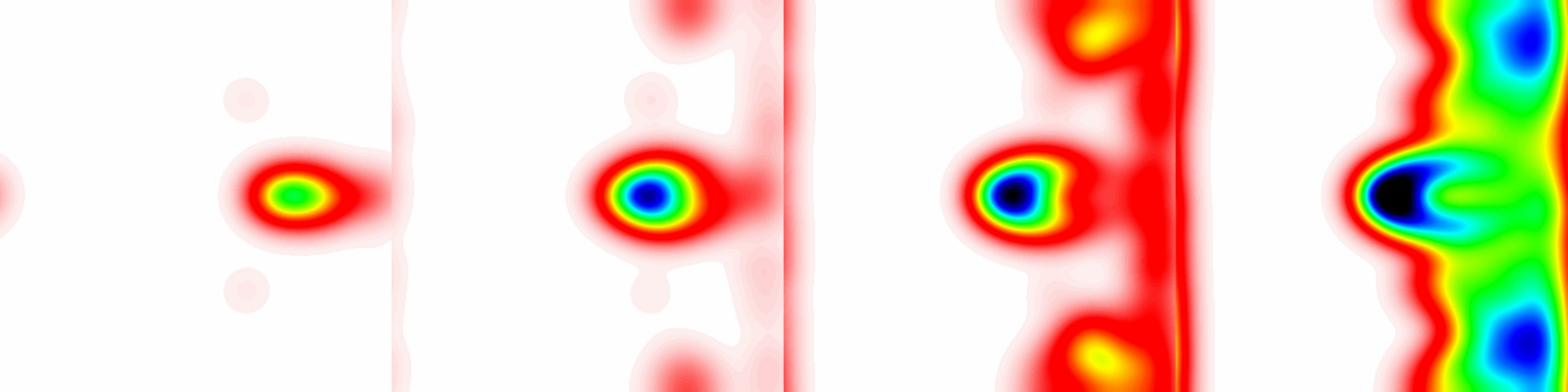}
\caption{A pseudocolour channel map of the medium flux model in $^{12}$CO. The channels, from left to right, are at $-7.02, -4.70, -2.27$ and $0.05$\,km\,s$^{-1}$. The colourbar represents the brightness temperature scale in Kelvin. Each channel is 4.87\,pc to a side.}
\label{med_cmap}
\end{figure*}

The medium flux $^{12}$CO images (Figure \ref{molImgs_med}) show a dense core at the tip of the cloud and a tail that gives the object a cometary appearance.  There is also fairly widespread emission about the cometary object which is due to foreground material, rather than material coinciding with the BRC itself. 
{A channel map of the medium flux model over the velocity range $-7.02$ to $+0.05$\,km\,s$^{-1}$ is given in Figure \ref{med_cmap}. We only include a channel map into the negative velocity range because there is no significant visual difference in the corresponding positive velocity channels. At low velocities (the right hand panel of Figure \ref{med_cmap}) most of the emission comes from the undisturbed material in the inner core at the tip of the cloud, as well as from the layers of the shell driving into the cloud perpendicular to the line of sight. At higher velocities the components of the shell driving into the shell along the line of sight dominate. At high negative velocities the shell from the near side of the cloud is observed and at high positive velocities the shell from the far side of the cloud is observed. }

The high flux $^{12}$CO images consist of a BRC which has a fairly dim bow compared to the low and medium flux models. The weaker extended emission behind the main cloud is therefore more easily visible due to the reduced contrast. There is also a lot of visible foreground material that is not directly associated with the BRC, towards the right of the image. The relative dimness {of the high flux model compared to the other two models} is because less material was accumulated during the high flux radiation hydrodynamic calculation in \cite{2012MNRAS.420..562H}. The $^{13}$CO and C$^{18}$O contours have fairly similar morphology, again tracing the $^{12}$CO extent of the gas well. 

In general, although the optically thin and thick lines may be focused on slightly different components of the cloud (depending on the density structure) each line traces a similar extent of the cloud for all models. It is therefore only necessary to use the combination of lines considered in this section to determine the average cloud conditions.

\subsection{Edge-on line profiles}
\label{kinematics}
We split {the edge--on image of each} cloud into a series of 20 equally sized boxes over which we calculate the average line profiles. These boxes are 50 by 50, 20 by 40 and 40 by 50 pixels for the low, medium and high flux models respectively. {The box sizes are chosen to provide optimal coverage of the BRC}. Signatures in profiles such as these are used to infer the kinematic behaviour of the gas by observers but the cause of these signatures is not always clear. {The interpretation of line profile features is usually based on simple theoretical models.} Since we have directly modelled the RDI process and know the thermal and kinematic conditions we can attempt to clarify the origin of some of these signatures. The 20 line profiles are shown for each model cloud across all three considered molecular species in Figure \ref{profiles}.

\subsubsection{General features of the edge-on line profiles}
At this viewing angle the line profiles typically consist of multiple components. The primary component is a peak of small width centred on $0$\,km\,s$^{-1}$ which is due to {turbulently} broadened emission from the stationary gas both within the cloud and from the back/foreground material. The secondary components are due to the swept up shell of material driving into the cloud, in agreement with findings from LOS velocity profiles calculated in \cite{2012A&A...538A..31T}. 

The profiles are all symmetric about the horizontal {mid--plane} of the BRC. There is no helical structure or apparent rotation of the cloud as is observed in some elephant trunks \citep{1998A&A...332L...5C,2003A&A...403..399C,2006A&A...454..201G}. {This result is unsurprising since the starting conditions of our RDI models were axisymmetric.} Rotating elephant trunks form via instability \citep{1980ApJ...240...84S,2011PASJ...63..795C} or the exposure of a turbulent medium to ionizing radiation \citep{2009ApJ...694L..26G,2010ApJ...723..971G,2012MNRAS.420..141E,2012A&A...546A..33T} rather than RDI. Unless the formation of trunks from a collection of initial inhomogeneities gives rise to a different velocity structure \citep{2010MNRAS.403..714M} it seems that the product of RDI of larger scale existing objects is the BRC, which is a distinct object from the narrower, relatively rapidly rotating elephant trunks. The velocities in most profiles are in a similar range to those identified in LOS velocity profiles of the RDI models of \cite{2010ApJ...723..971G}. Some of the cloud-averaged profiles from \cite{2009A&A...497..789U} also show secondary features that span the velocity range illustrated here, for example SFO81, which shows two secondary peaks separated by around 9\,km\,s$^{-1}$. {The range in velocities between the peaks in the line profiles observed in \cite{2009A&A...497..789U} are too large to be the result of self absorption. In our the models this large velocity range is due to the systematic bulk shell motions relative to the low velocity gas encompassed by the shell. }

\begin{figure*}
	\hspace{-10pt}
	\includegraphics[width=4.5cm]{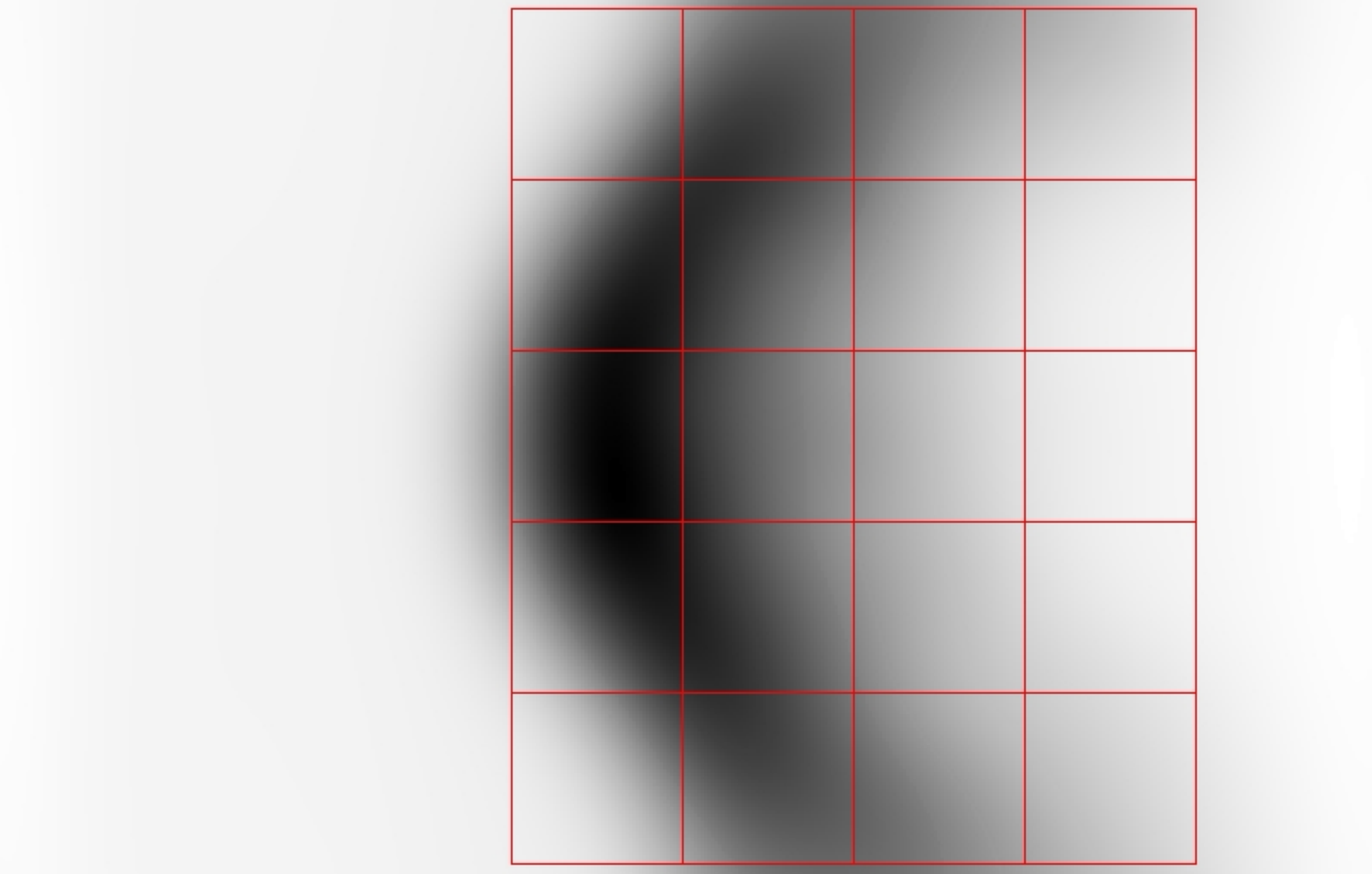}
	\hspace{40pt}
	\includegraphics[width=4.5cm]{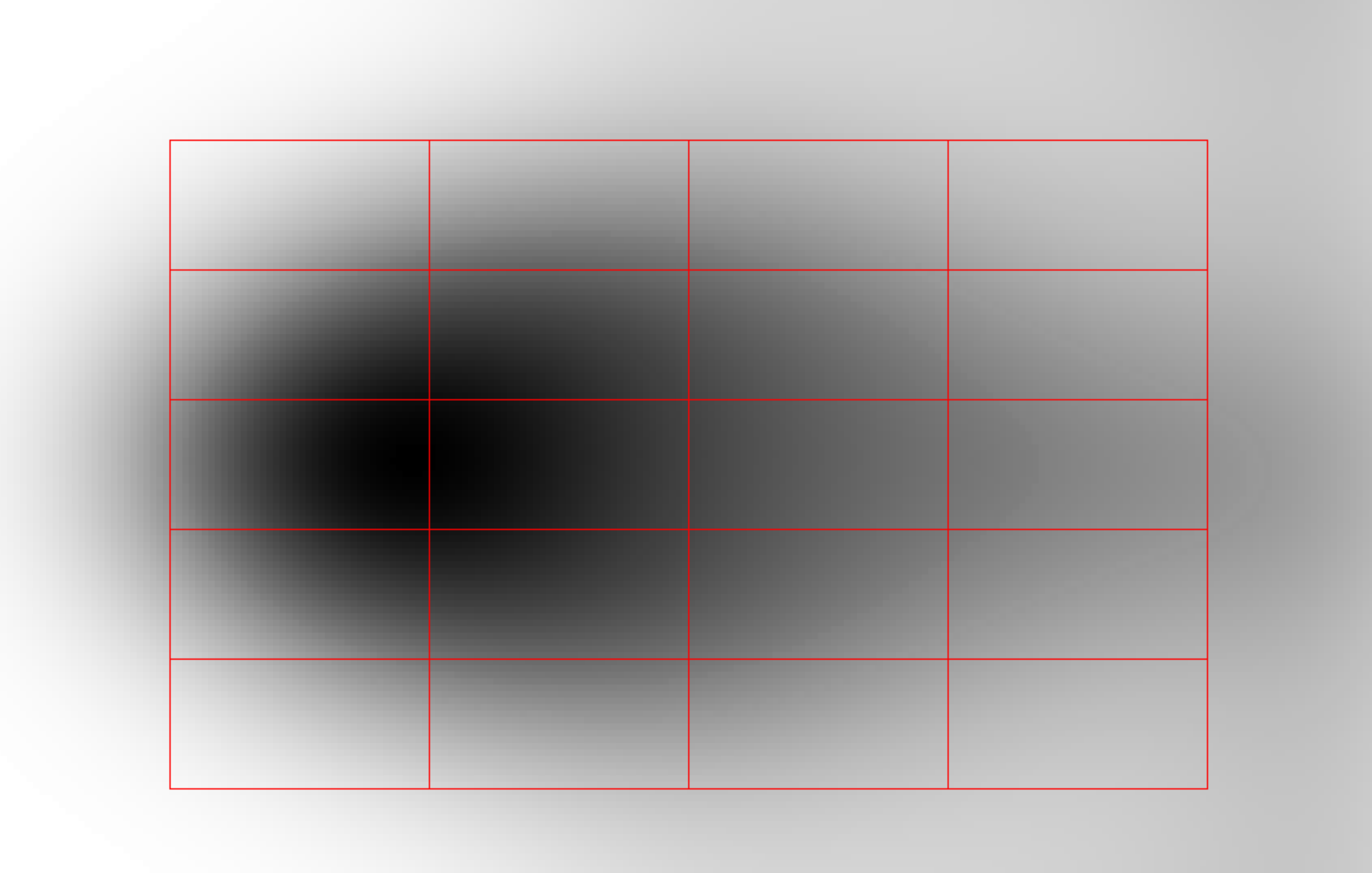}
	\hspace{20pt}
	\includegraphics[width=4.5cm]{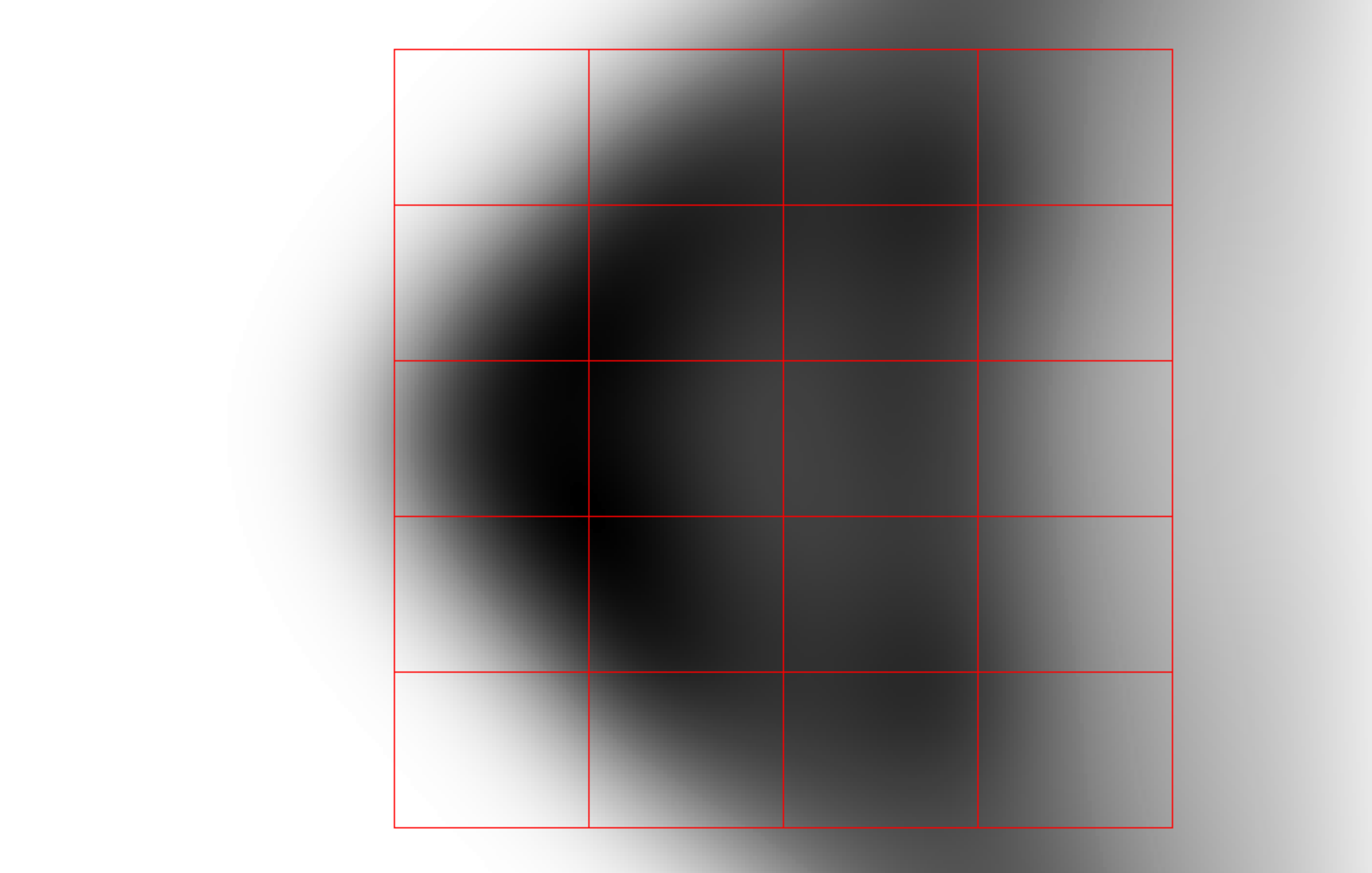}

	\includegraphics[width=5.4cm]{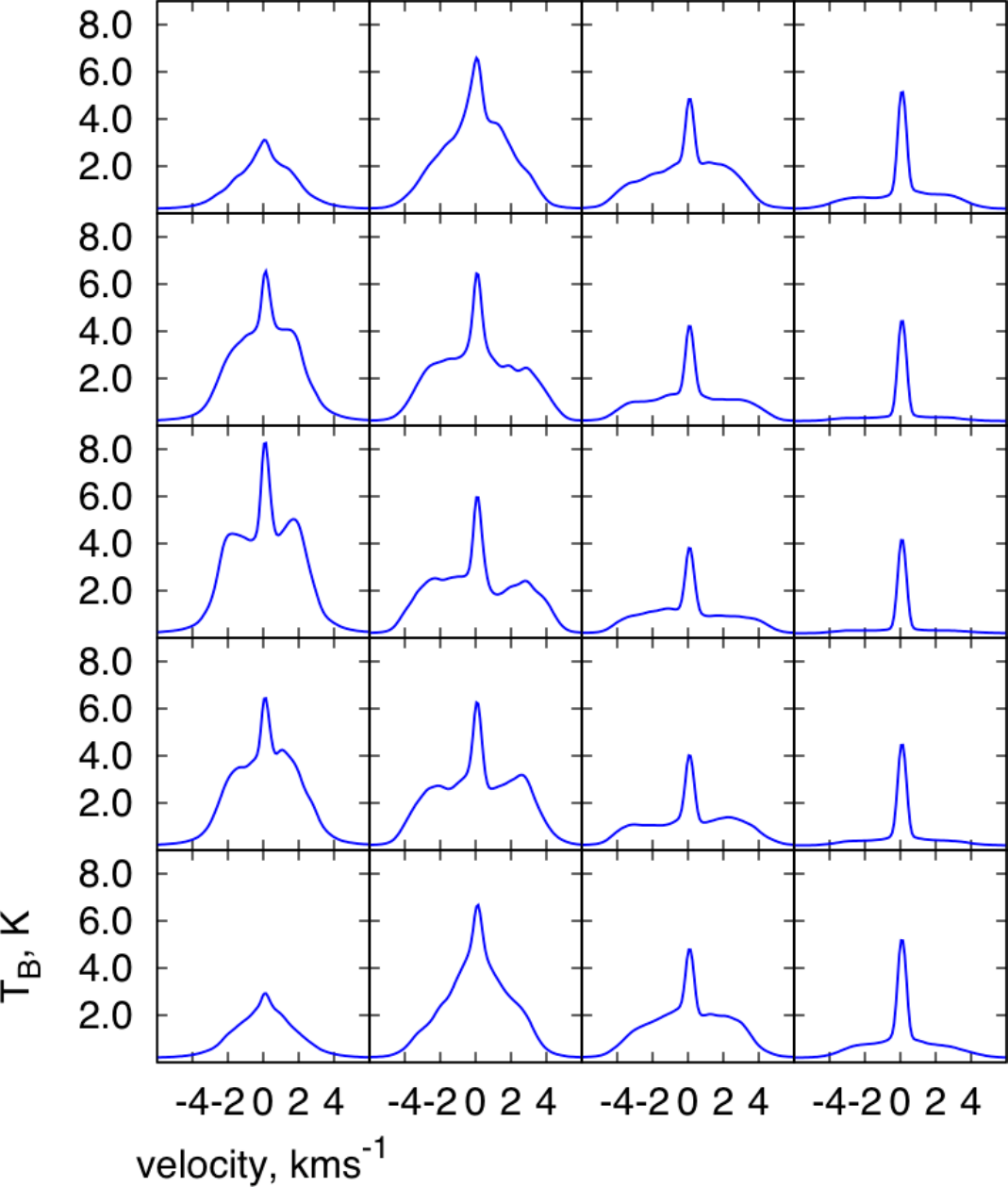}
	\includegraphics[width=5.4cm]{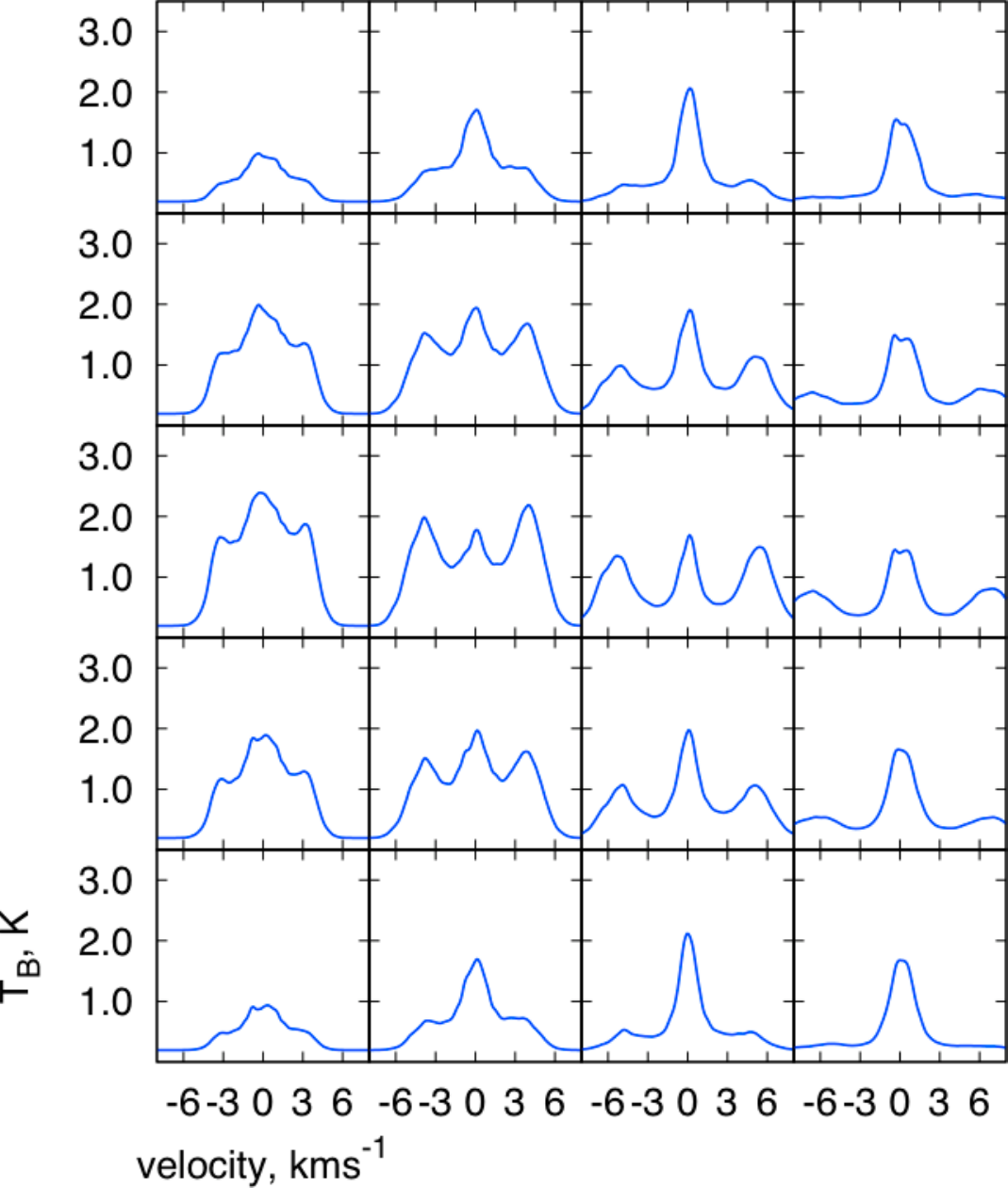}
	\includegraphics[width=5.4cm]{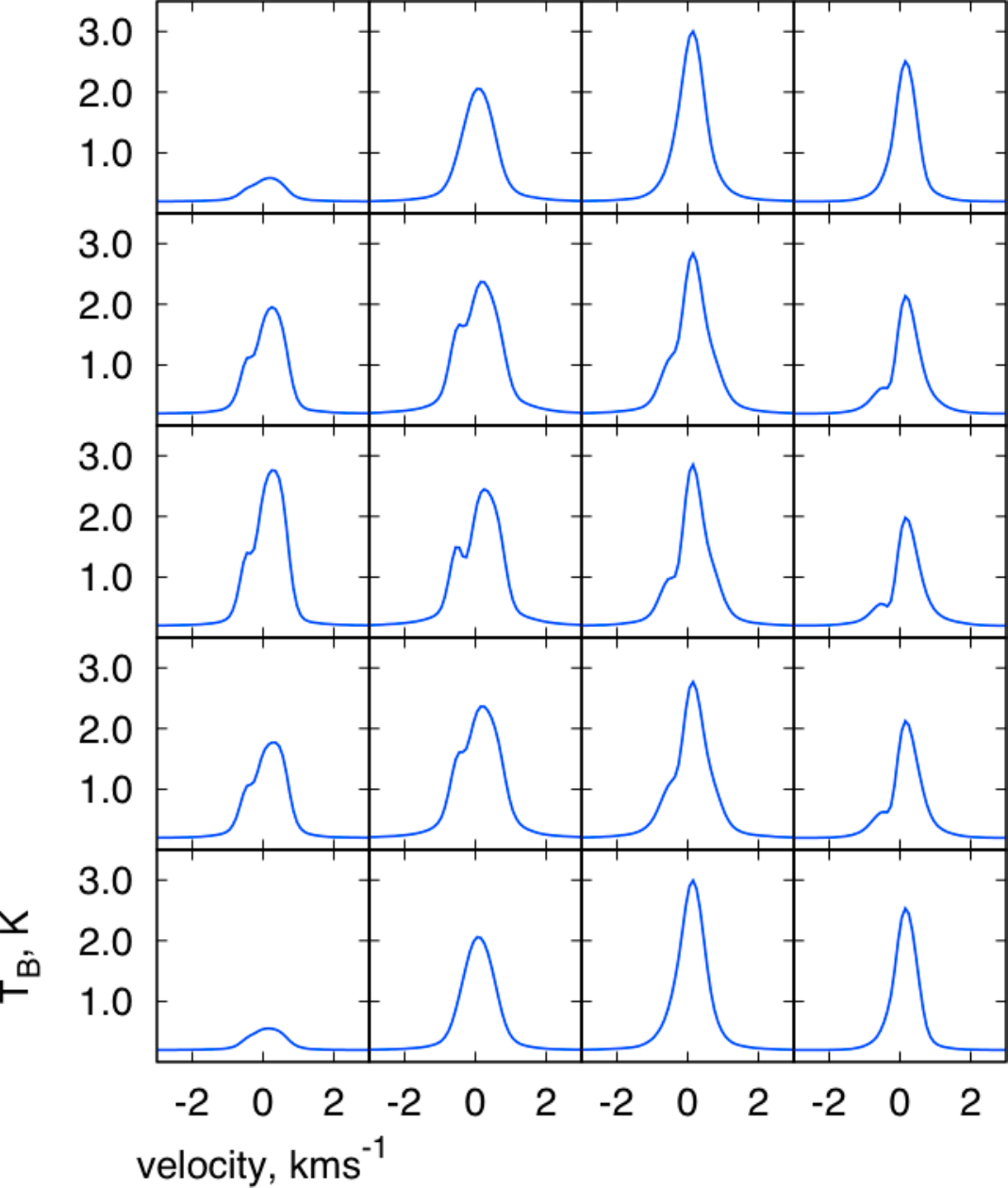}

	\includegraphics[width=5.4cm]{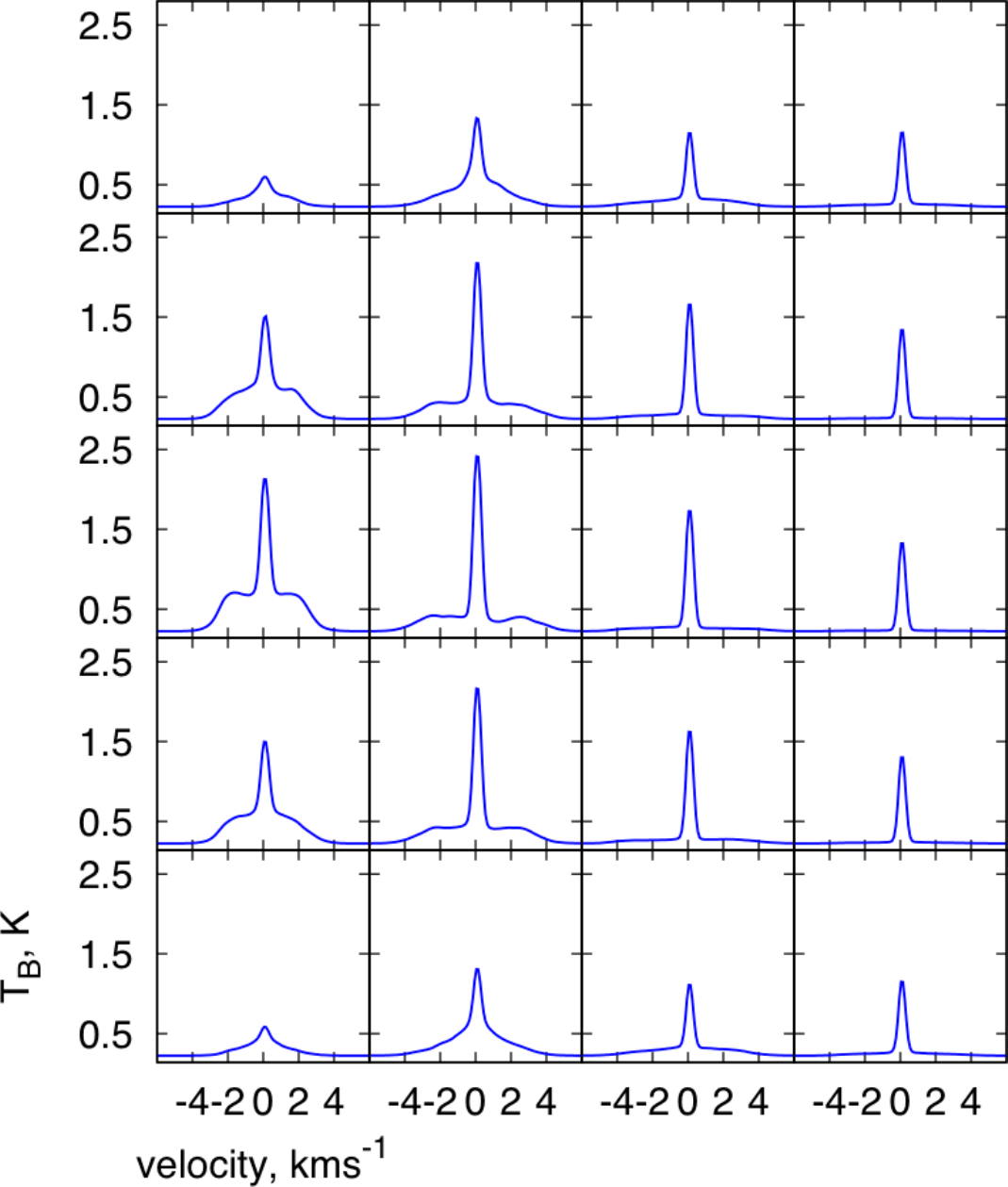}
	\includegraphics[width=5.4cm]{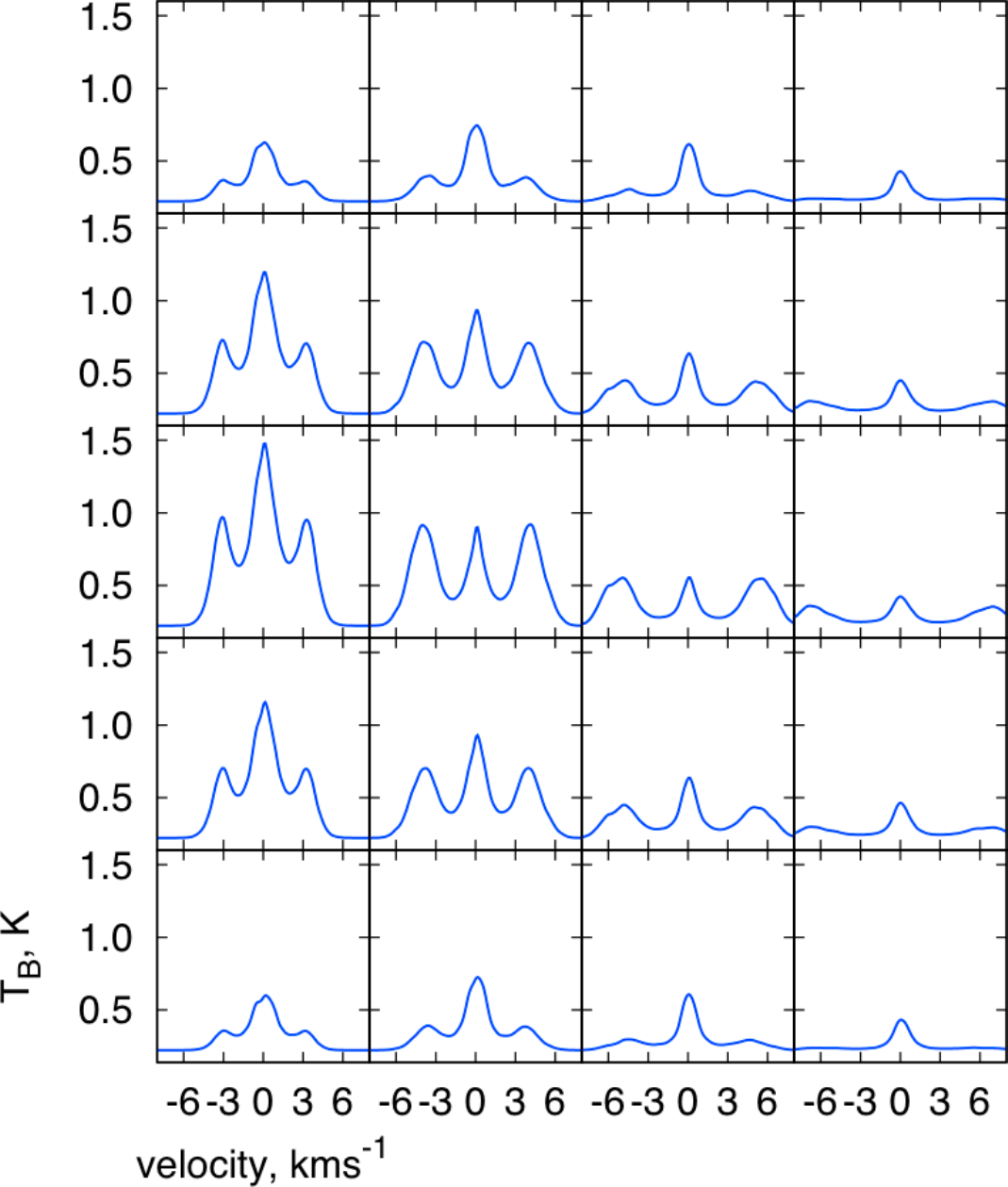}
	\includegraphics[width=5.4cm]{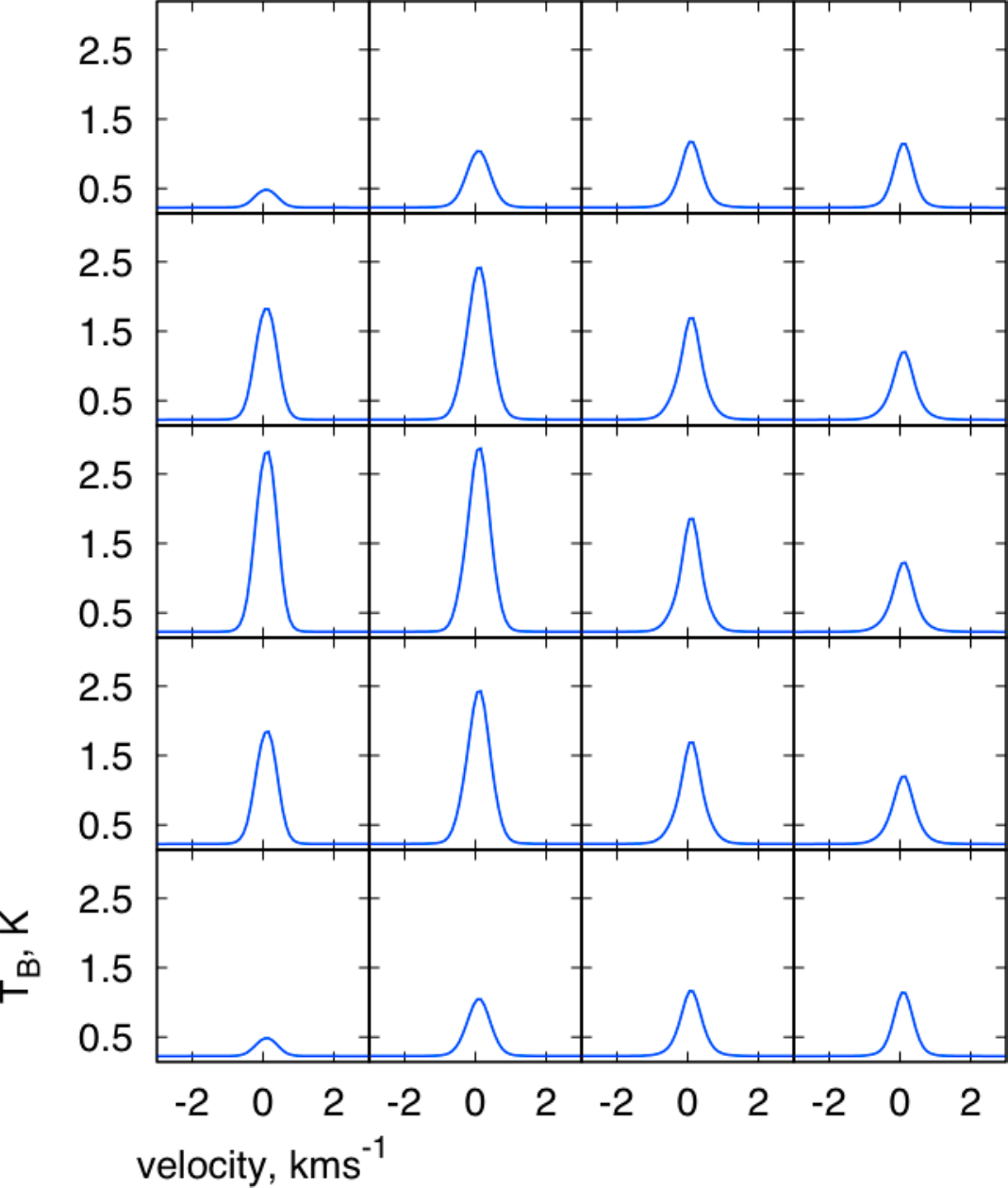}

	\includegraphics[width=5.4cm]{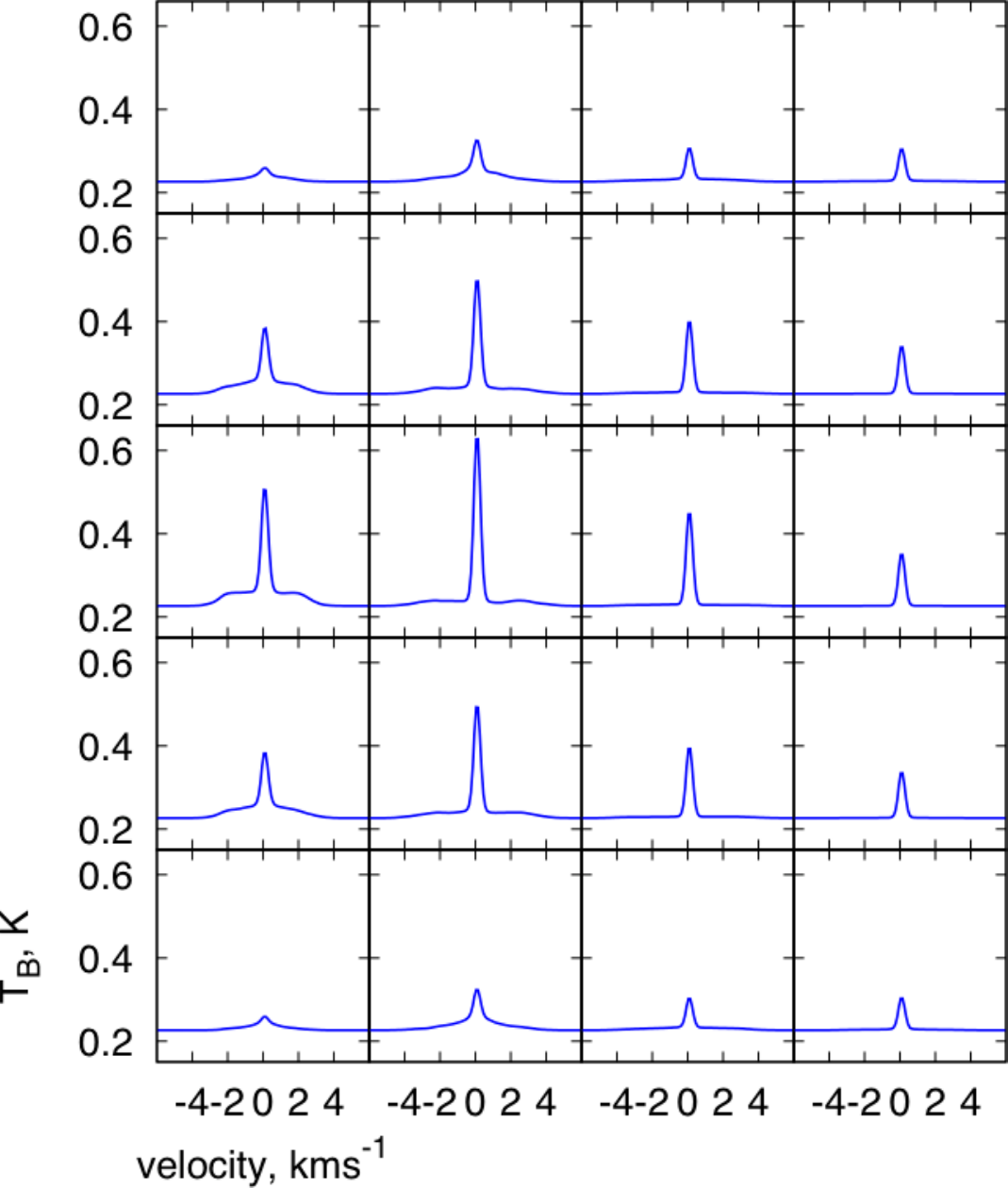}
	\includegraphics[width=5.4cm]{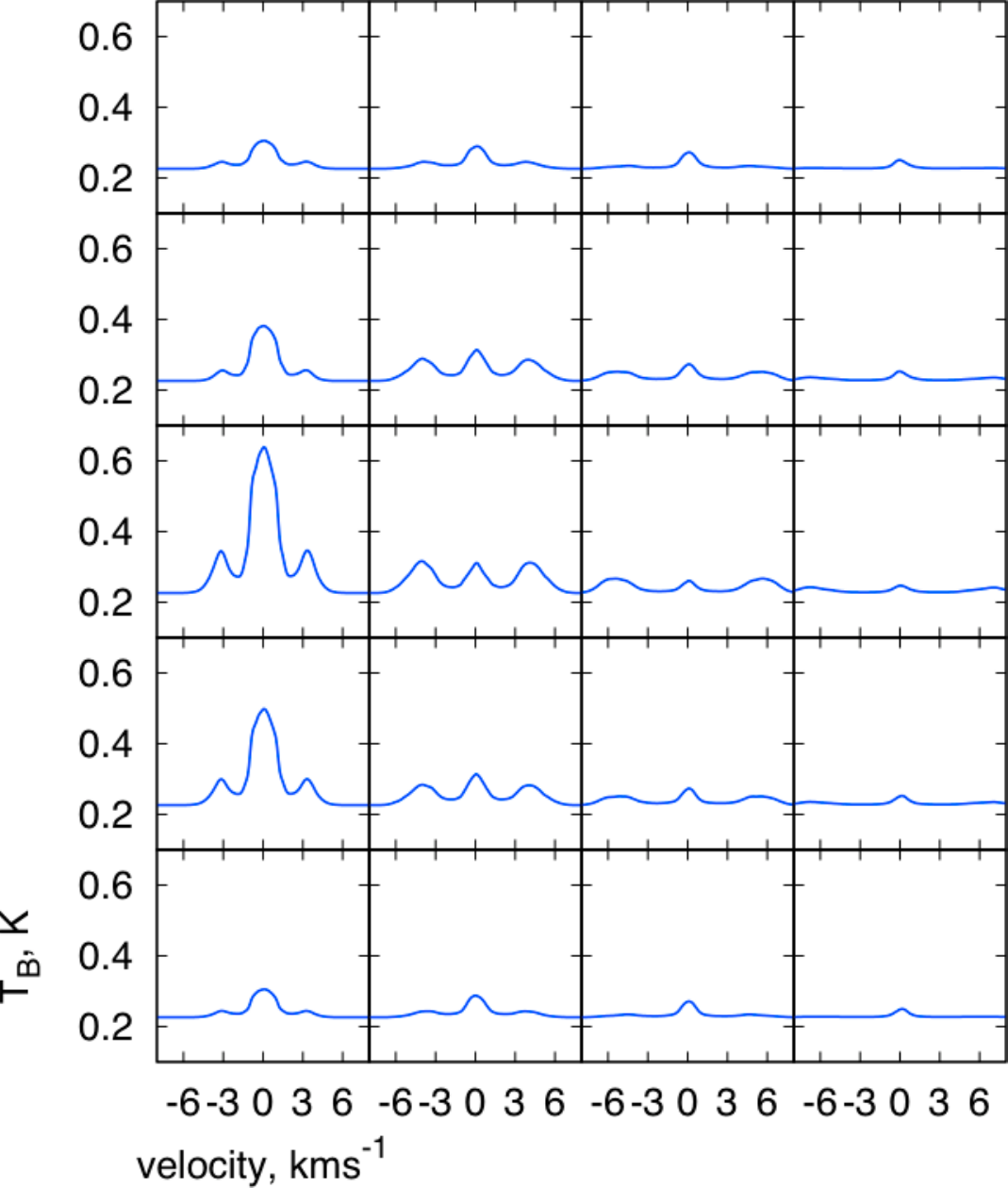}
	\includegraphics[width=5.4cm]{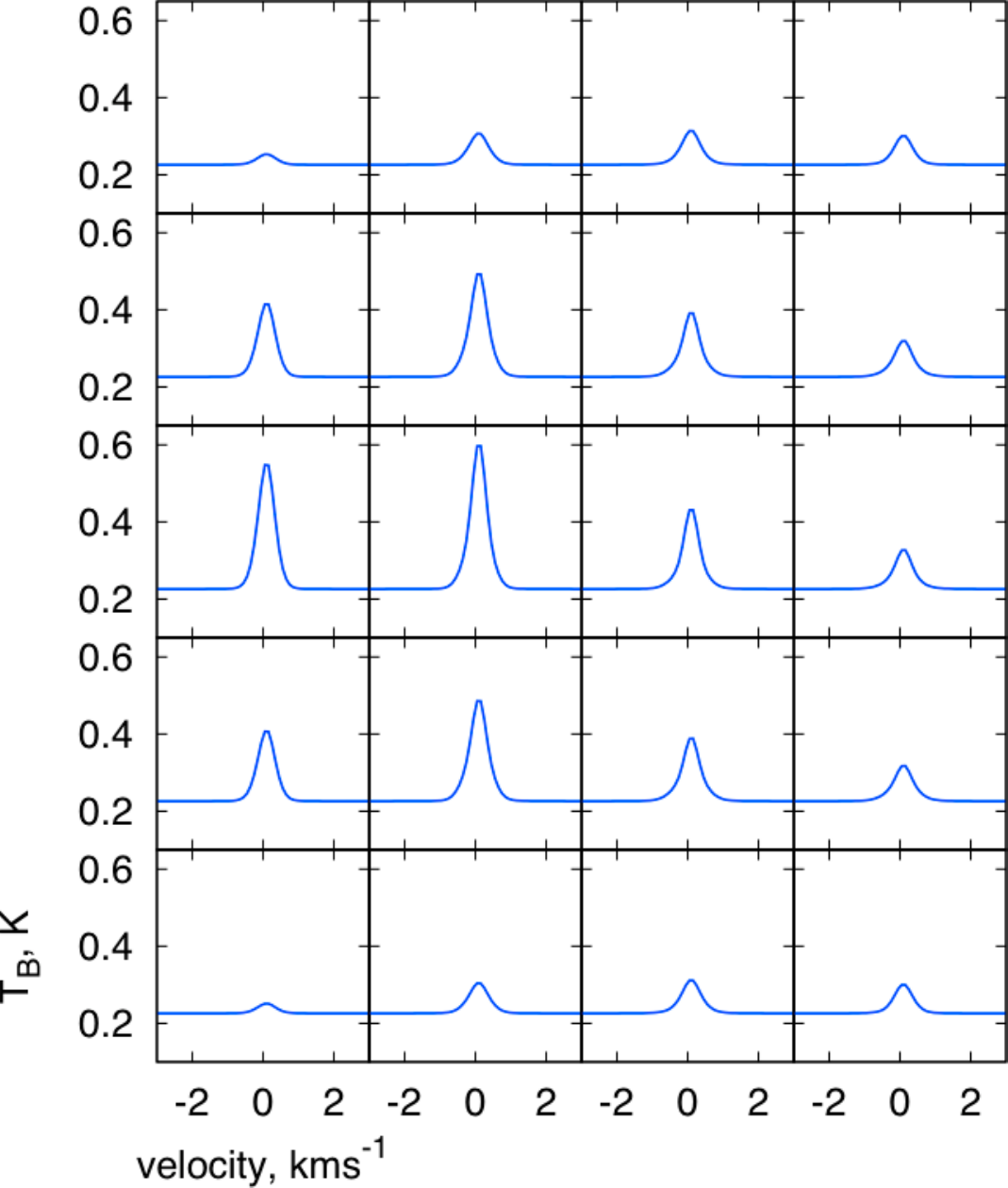}
	\caption{Line profiles for regions of the low, medium and high flux models from left to right columns respectively. The top row shows the region locations. The profiles are, from top to bottom, $^{12}$CO, $^{13}$CO and C$^{18}$O.}
	\label{profiles}
\end{figure*}

{Features in the line profiles due to the shell have their intensity determined by the density at which the line becomes optically thick. The peak velocity of the shell feature depends on the viewing angle of the observer and the propagation direction of the shell. For example if the shell is travelling perpendicularly to the observer viewing angle then the velocity of the shell peak will be slower than if the observer is viewing the shell along its propagation vector.} 
In the low flux case the shell layers give rise to broad shoulders about the central peak {of relatively low intensity compared to the central peak.} 
{In the medium flux case} the {shell contributions to the line profile} manifest themselves as separate peaks, since the shell is sufficiently dense and travelling sufficiently fast for the secondary peaks to be isolated from the {turbulently} broadened emission of the stationary cloud. 
In the high flux model the shell layer is thinner and propagating more slowly so the distinction between the uncompressed cloud and shell is not as clear. {The peaks due to the shell in these models increase and decrease in strength as the observer moves to different viewing angles, as discussed further in section \ref{inc}. }

The {position of the boxes with the most intense line profile peak} in the low and high flux models in Figure \ref{profiles} {do} not correlate between $^{12}$CO and C$^{18}$O. In the medium flux model, the {position of the most intense $^{12}$CO and C$^{18}$O profile peaks} do correlate, being situated towards the tip of the cometary object in the left most column of the middle row. {This implies that in the high and low flux model the optically thick and thin lines are predominantly probing different parts of the clouds. In the medium flux case the optically thick and thin line both probe the same part of the cloud. }

\subsubsection{Edge-on asymmetries}
Another interesting feature is that asymmetries in the $^{12}$CO line profiles at this edge-on viewing angle are predominantly red. That is, non-Gaussian features with $v > 0$\,km\,s$^{-1}$ are stronger than those with $v < 0$\,km\,s$^{-1}$. {An example of red asymmetry from Figure \ref{profiles} is the central row of the medium flux model in $^{12}$CO where the shells are most directly propagating towards and away from the observer.} The reason for this red-asymmetry is the dense shell of material that is driving into the cloud. The BRC observed from this viewing angle is a three-component system, with a central (and ambient) gas cloud, a near-shell propagating away from the observer into the cloud (red-shifted) and a far-shell propagating towards the observer into the cloud (blue-shifted). The optically thick emission from each component will only be from the closest layer to the observer at a given velocity. The interior edge of the driving shell is typically at a lower density than the exterior and central regions of the shell so the blue peak in the line profile is from lower density interior gas and is therefore weaker. At the frequency of the CO molecular transitions considered here, dust absorption plays a negligible role in attenuating the observed intensity from the far shell layer.
 Asymmetries become more pronounced at different viewing angles. For example, as the observer moves to view the BRC face on the shell will be denser and moving more directly along the observer's line of sight and the profile will be more red-asymmetric. This is discussed  further in section \ref{inc}.

\subsubsection{Envelope expansion with core collapse}
\begin{figure}
	\hspace{-15pt}
	\includegraphics[width=9.8cm]{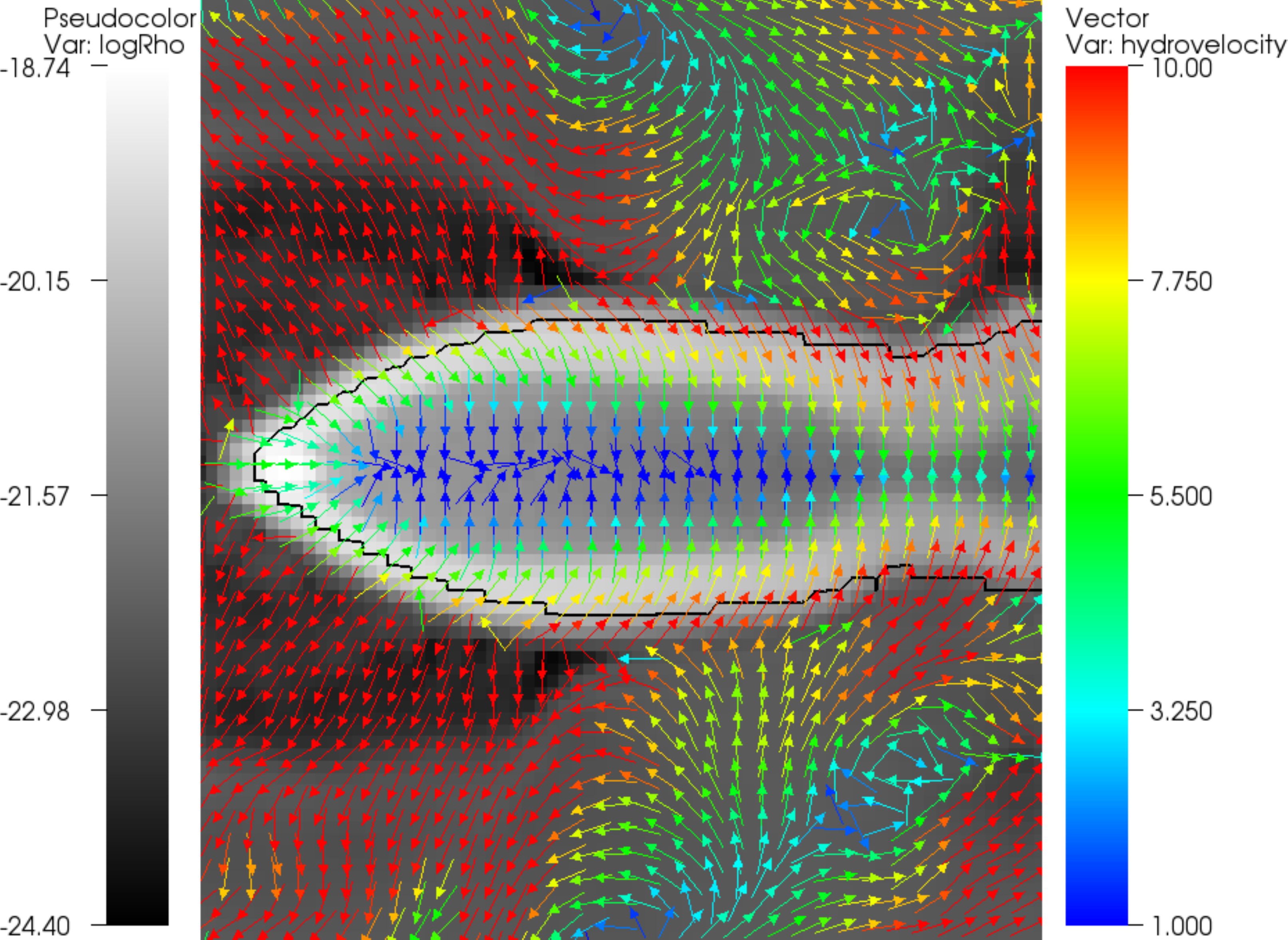}
	\caption{A slice through the logarithmic density distribution of the medium flux model. Overlaid are velocity vectors and a contour corresponding to a neutral atomic hydrogen fraction of 0.1. Material external to this contour around the cloud will not contribute to the molecular line data cubes because the gas is ionized. Note that the vortices in the hot gas are Kelvin-Helmholtz instabilities resulting from shear between the driving flow and photo-evaporative outflow. The greyscale bar is the logarithmic density in $\log_{10}$(g\,cm\,$^{-3}$) and the colour bar is for the velocity vectors in km\,s$^{-1}$. This cut is 3.9 by 2.6\,pc.}
	\label{medwvecs}
\end{figure}

An alternative explanation for the red-asymmetry {in BRC line profiles} is the envelope expansion with core collapse (EECC) model, in which the cloud is a two component system with a collapsing core and an expanding outer shell, \citep[e.g.][]{2006ApJ...652.1366K,2010MNRAS.403.1919G,2011MNRAS.412.1755L,2011ApJ...741..113F}. Figure \ref{medwvecs} shows a slice through the medium flux model density distribution and has a contour corresponding to the neutral atomic hydrogen fraction being equal to 0.1 overlaid. This figure clearly illustrates that the gas outflowing towards the observer is all too ionized to harbour molecular gas {(see section \ref{num_meth})}. The contributor to the line profile must be the neutral part of the driving shell, the gas contained within it and any neutral foreground material. The EECC models of \cite{2010MNRAS.403.1919G}, \cite{2011MNRAS.412.1755L} and \cite{2011ApJ...741..113F} {describe well smaller isolated starless cores that are not being driven by the surroundings but following our result} probably do not extend to BRCs and RDI, {where the high velocity motions of the dense shell dominate the line profile.}

\subsubsection{Comparison of line profiles with observations}
SFO 81 is a BRC studied in \cite{2009A&A...497..789U} and has a  triple-peaked {$^{12}$CO} profile suggesting it may be viewed edge on. Obtaining line profiles over smaller regions of this cloud, in the manner of this section, would help to confirm this. 

\cite{2009A&A...497..789U} also {present} a number of other profiles that have similarities to those here, SFO 59, 60, 73, 80, 81, 86 and 87 are all multi-peaked in $^{12}$CO. Interestingly, \cite{2009A&A...497..789U} suggest that SFO 80, 81 and 86 are all unlikely to be triggered because a PDR is not readily observed at $8\,\mu\rm{m}$ {\citep[these observations were primarily made using the Midcourse Space Experiment satellite,][]{2001AJ....121.2819P}}. SFO 73 and 87, however, do have a visible PDR at $8\,\mu\rm{m}$ and are expected to be sites of triggering. What the profiles of the apparently un--triggered clouds (SFO 80, 81 and 86) have in common, compared to SFO 87, is that the stronger (or only visible) of the secondary peaks is blue shifted (the SFO 73 {line} profile is too complex to {compare with the others}).  If these blue shifted secondary peaks are due to the shell, then our results imply that the observer is viewing the cloud from behind with the shell moving towards them. As such it is {less} surprising that the PDR is not so readily visible, as it would be on the opposite side of the {(potentially optically thick)} cloud to the observer. SFO 87 has a strong secondary red peak, suggesting the shell is moving away from the observer and the cloud is being viewed face on. This is supported by the fact that the PDR is readily visible for SFO 87. {We conclude that} not viewing a substantial PDR at shorter wavelengths may not be sufficient to rule out triggering in a BRC.
Follow up analysis with longer wavelength {Herschel or} Spitzer archival data could help to identify a PDR in the BRCs where one was not identified at $8\,\mu\rm{m}$.

There {are} also a number of wings and shoulders identified {in the line profiles} given in \cite{2009A&A...497..789U}, such as SFO 51, 55, 71 and 79 that resemble the features of the low flux and high flux models at this inclination and all models at higher inclinations.

It should be noted that although these {edge--on} profiles best illustrate the various contributing components of the BRC, the form of a profile changes rapidly with viewing angle. For example, the three strong peaks of the medium flux profile will be dominated by a single peak due to the shell with non--Gaussian wings as the observer moves in front or behind of the BRC. That an edge-on viewing angle is comparatively rare is the reason that single peaked profiles tend to occur more frequently {in observations to date} \citep{2009MNRAS.400.1726M,2009A&A...497..789U}.

\subsection{The effect of viewing angle on line profiles}
\label{inc}
\begin{figure}
	\includegraphics[width=8.5cm]{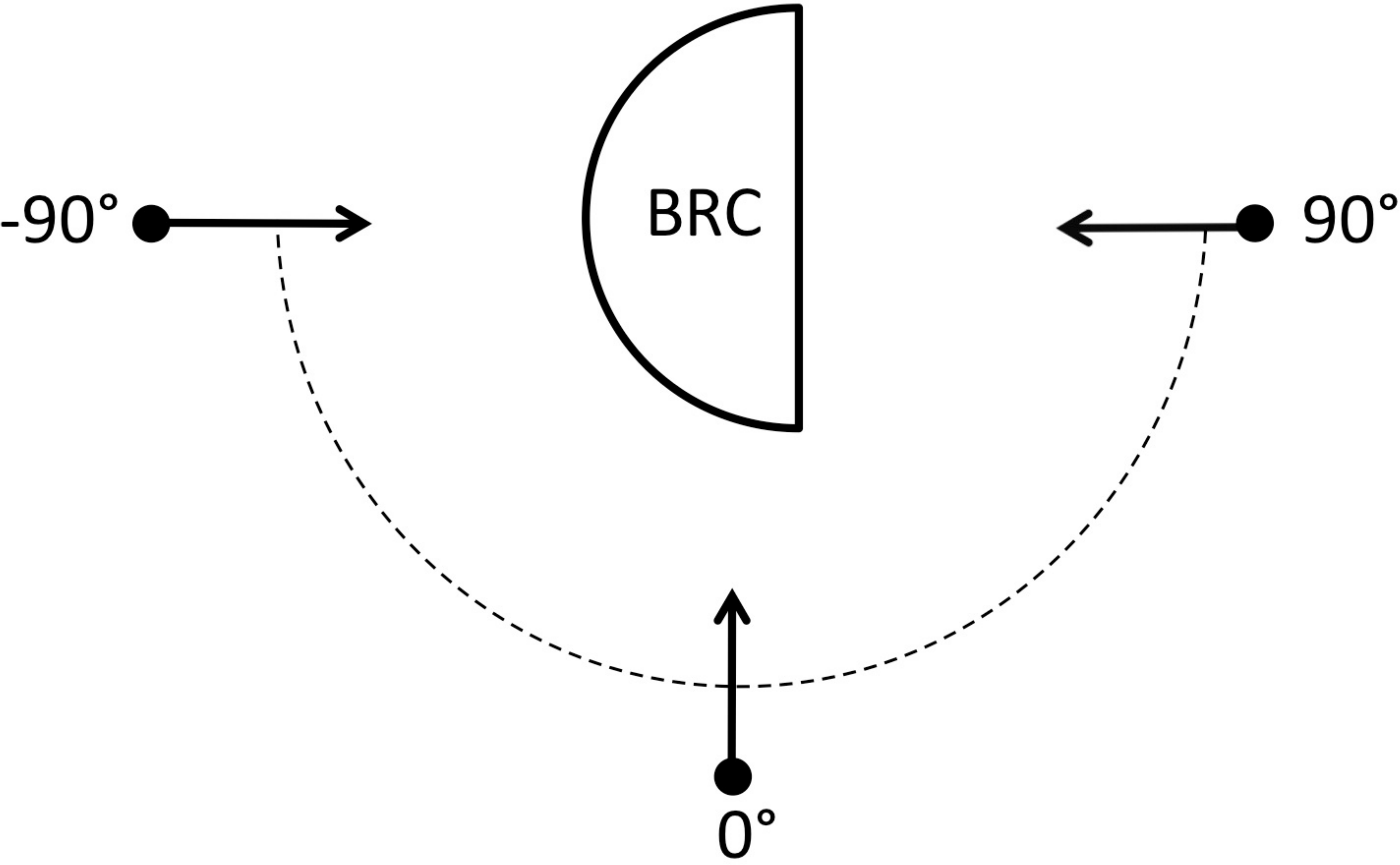}
	\caption{A schematic of the viewing angle convention used. At $-90^{\circ}$ the observer is facing the bright rim of the class A cloud. At $90^{\circ}$ the observer is behind the cloud. }
	\label{schem}
\end{figure}

We generated data cubes for $^{12}$CO and C$^{18}$O from $-90$ degrees (face on to the BRC) to $90$ degrees (behind the BRC) in intervals of $15$ degrees. A schematic of these viewing angles is given in Figure \ref{schem}. Due to the large volume of data it is impractical to replicate the overlaid grid analysis presented for the edge on viewing angle in section \ref{kinematics} for each inclination. We therefore focus on the overall variation in the {average} line profile {over the cloud}. 

\subsubsection{The variation of the line profile peak intensity velocity}
\begin{figure}
	\hspace{-5pt}
	\includegraphics[width=8.6cm]{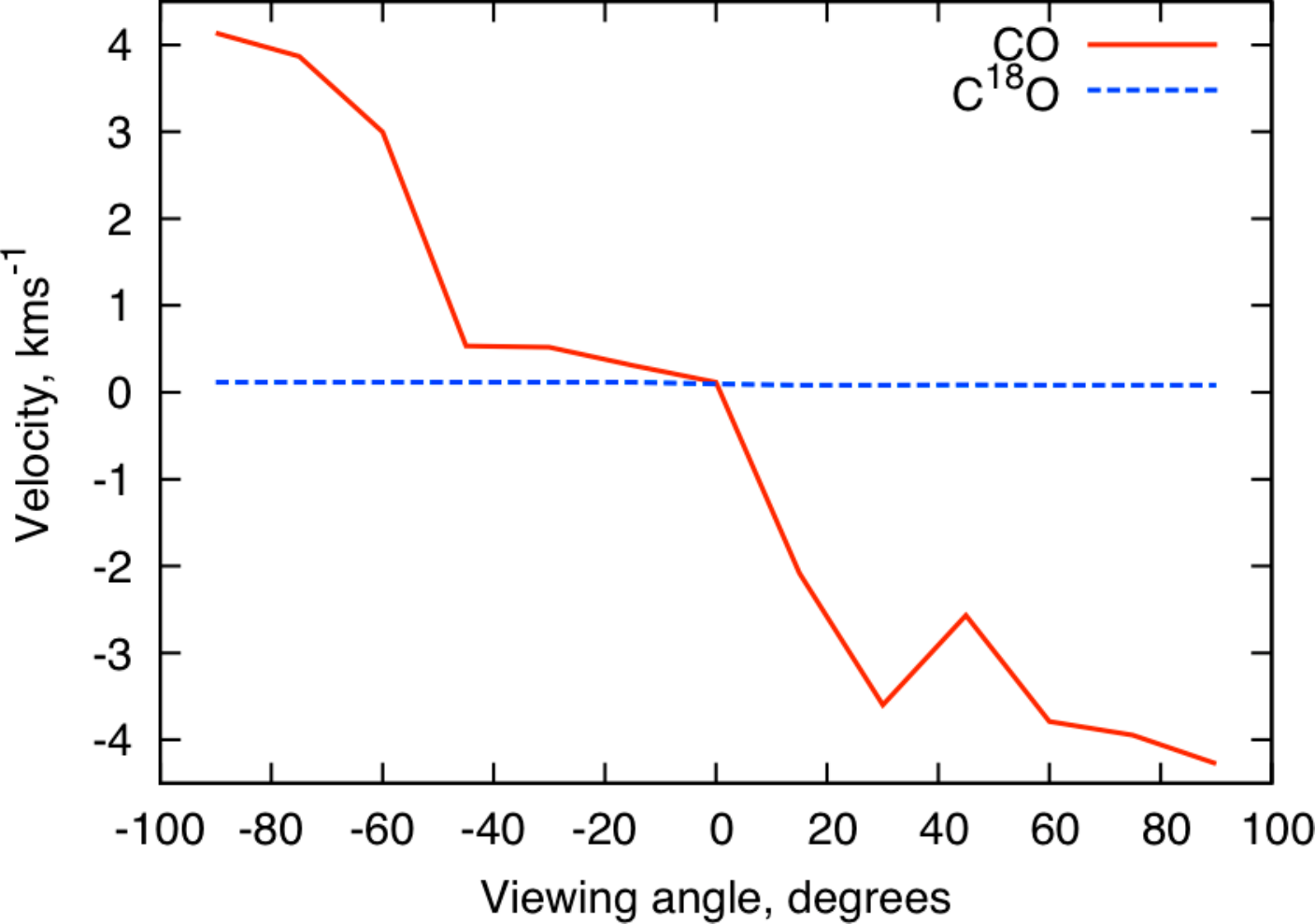}

	\vspace{15pt}
	\hspace{-5pt}
	\includegraphics[width=8.6cm]{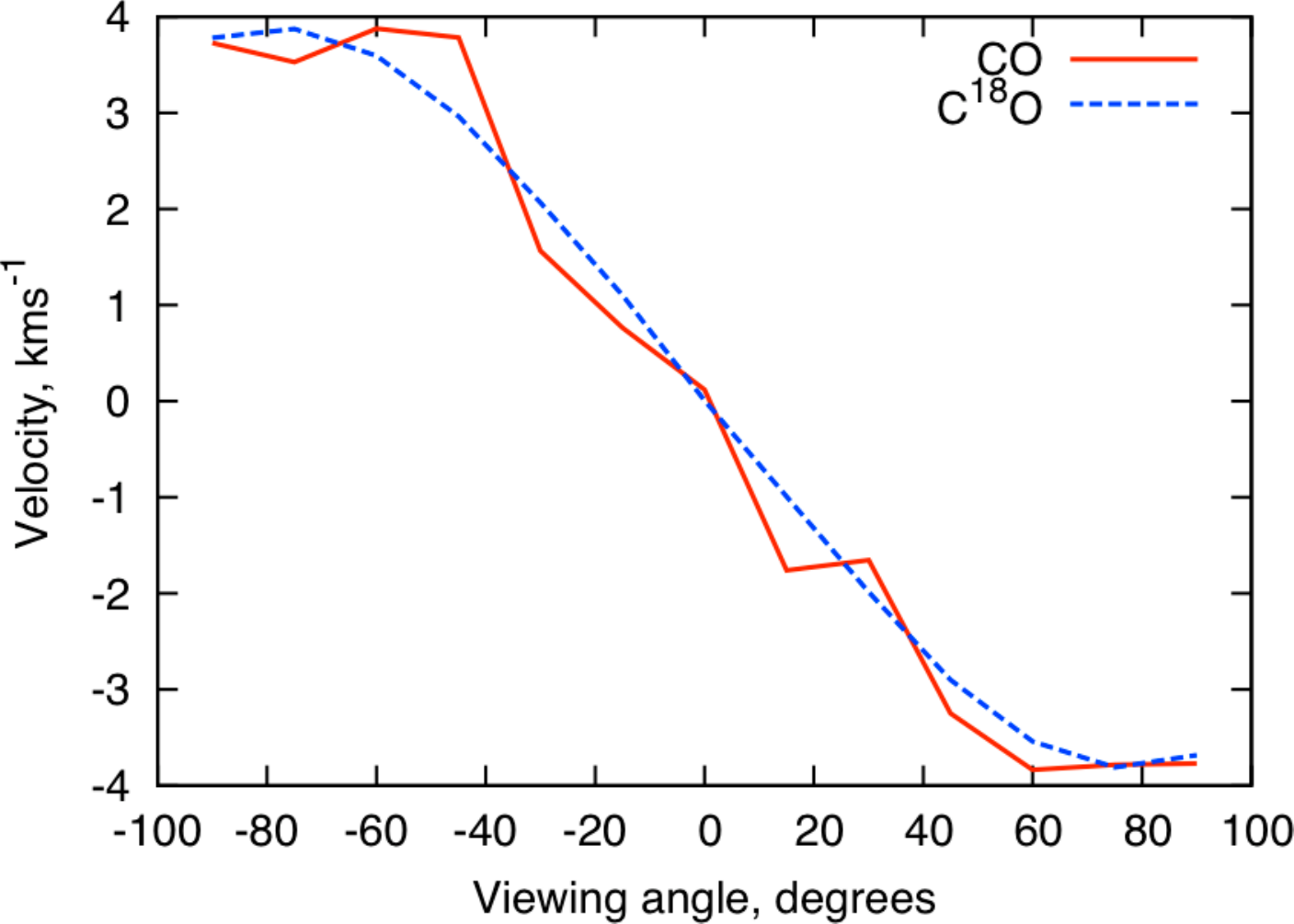}

	\vspace{15pt}
	\hspace{-15.5pt}
	\includegraphics[width=9.cm]{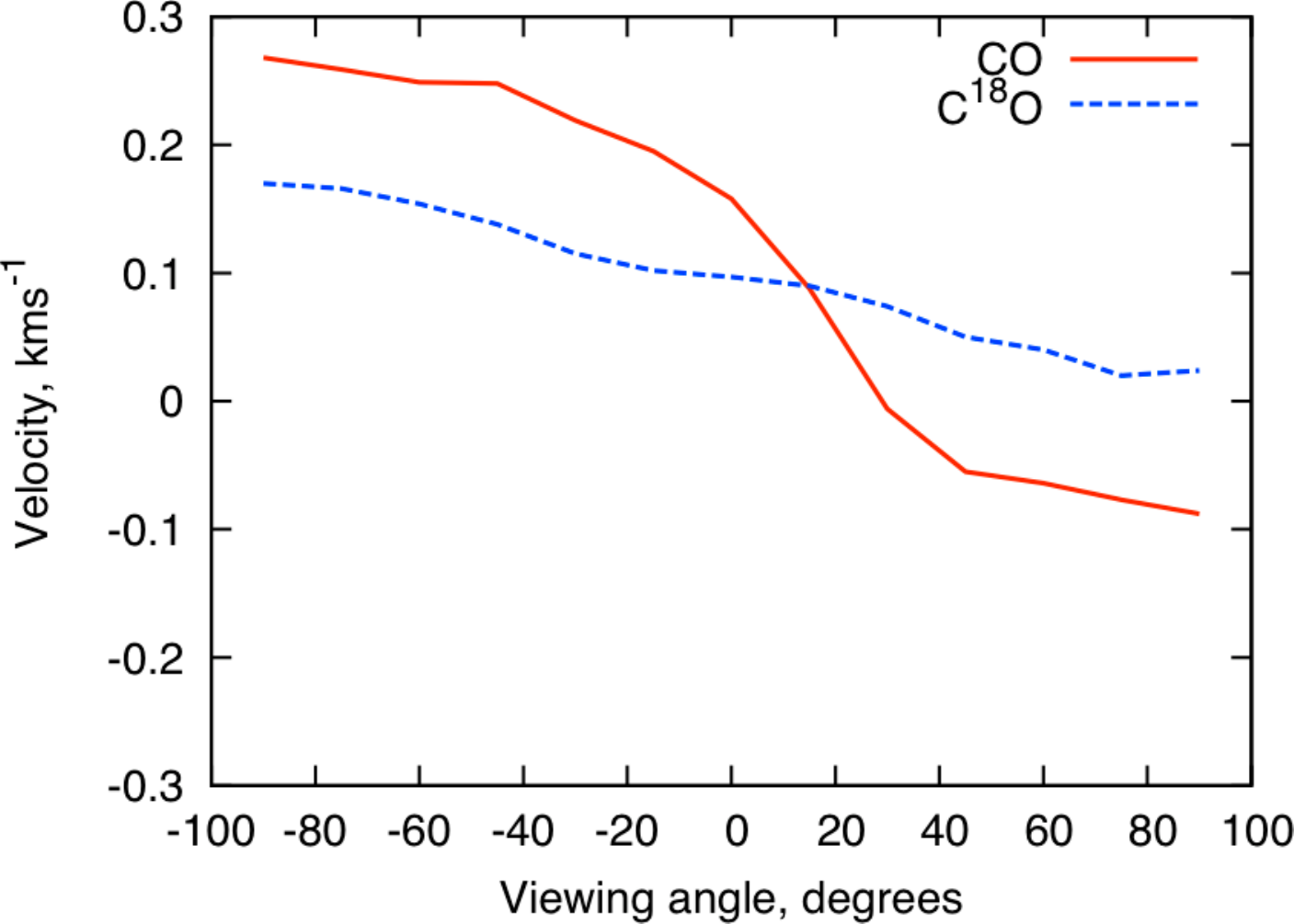}
	\caption{The variation of the {velocity at which the average line profile over the cloud is at maximum intensity} with viewing angle for the low (top), medium (middle) and high (bottom) flux models.} 
	\label{VvsI}
\end{figure}

We {plotted} the {velocity at which the average line profile over the cloud is at maximum intensity (hereafter referred to as velocity for brevity)} as a function of viewing angle for {both the $^{12}$CO and C$^{18}$O} lines across all models in Figure \ref{VvsI}. 

\begin{figure}
	\includegraphics[width=8cm]{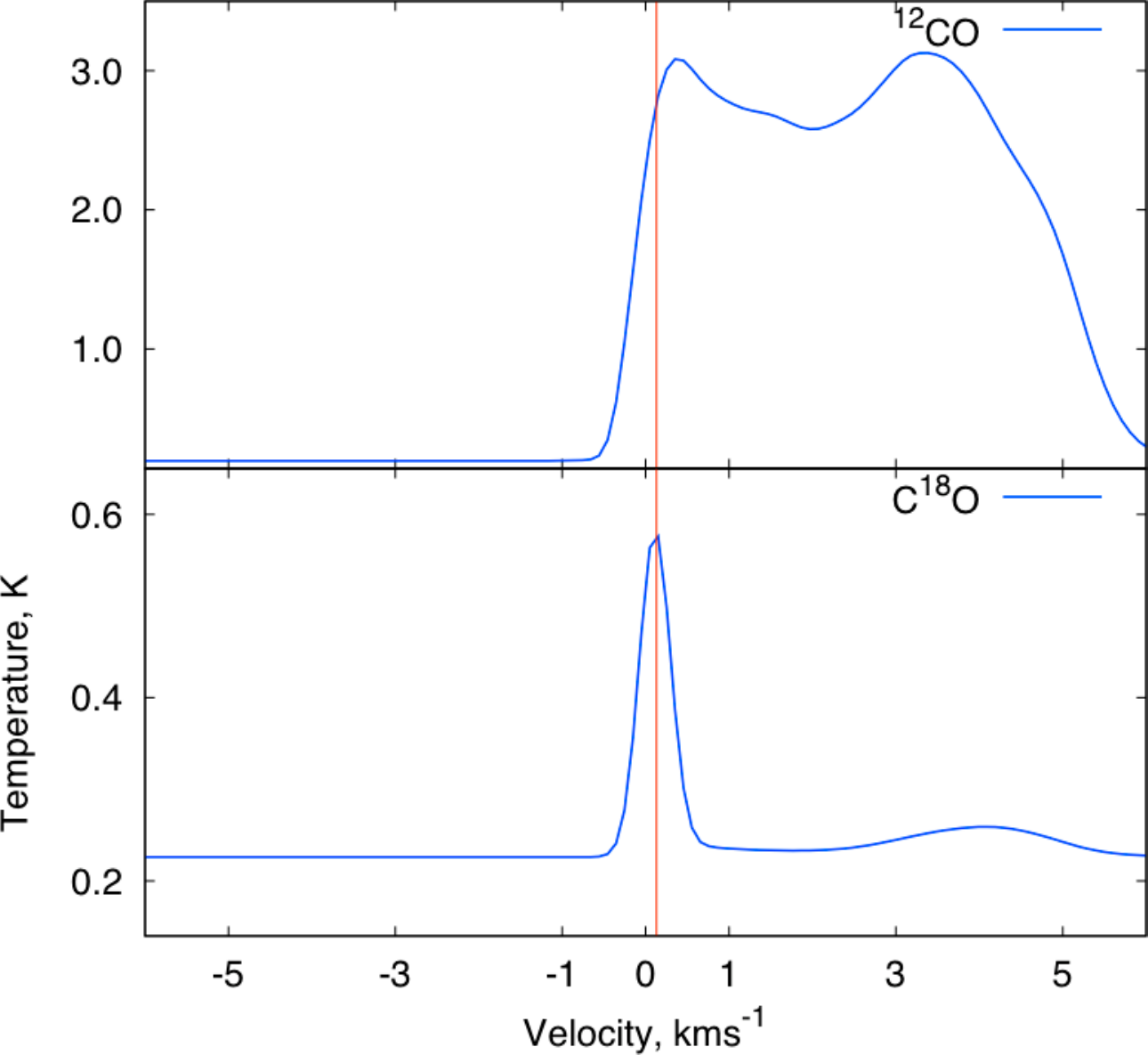}
	\includegraphics[width=8cm]{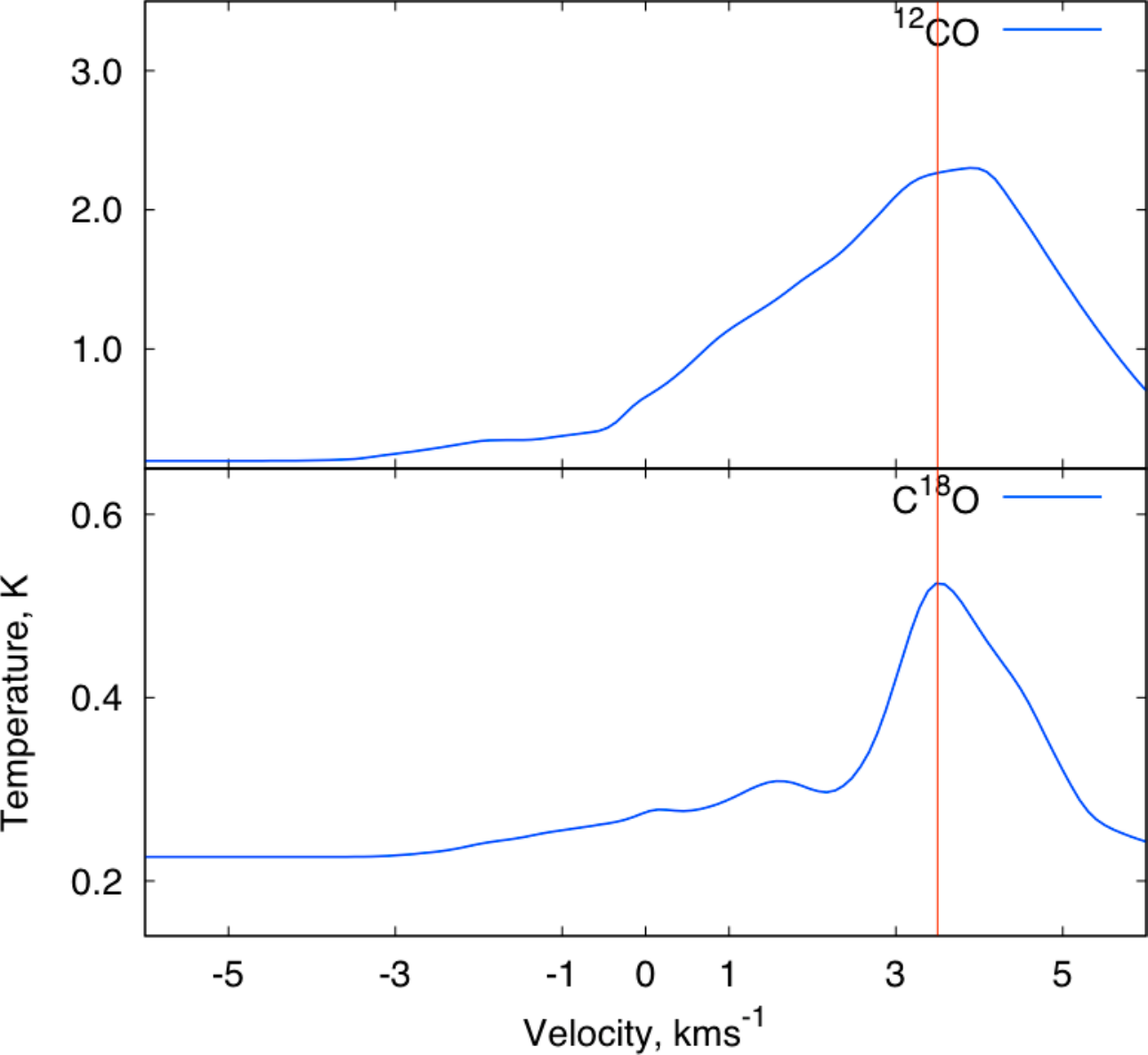}
	\includegraphics[width=8cm]{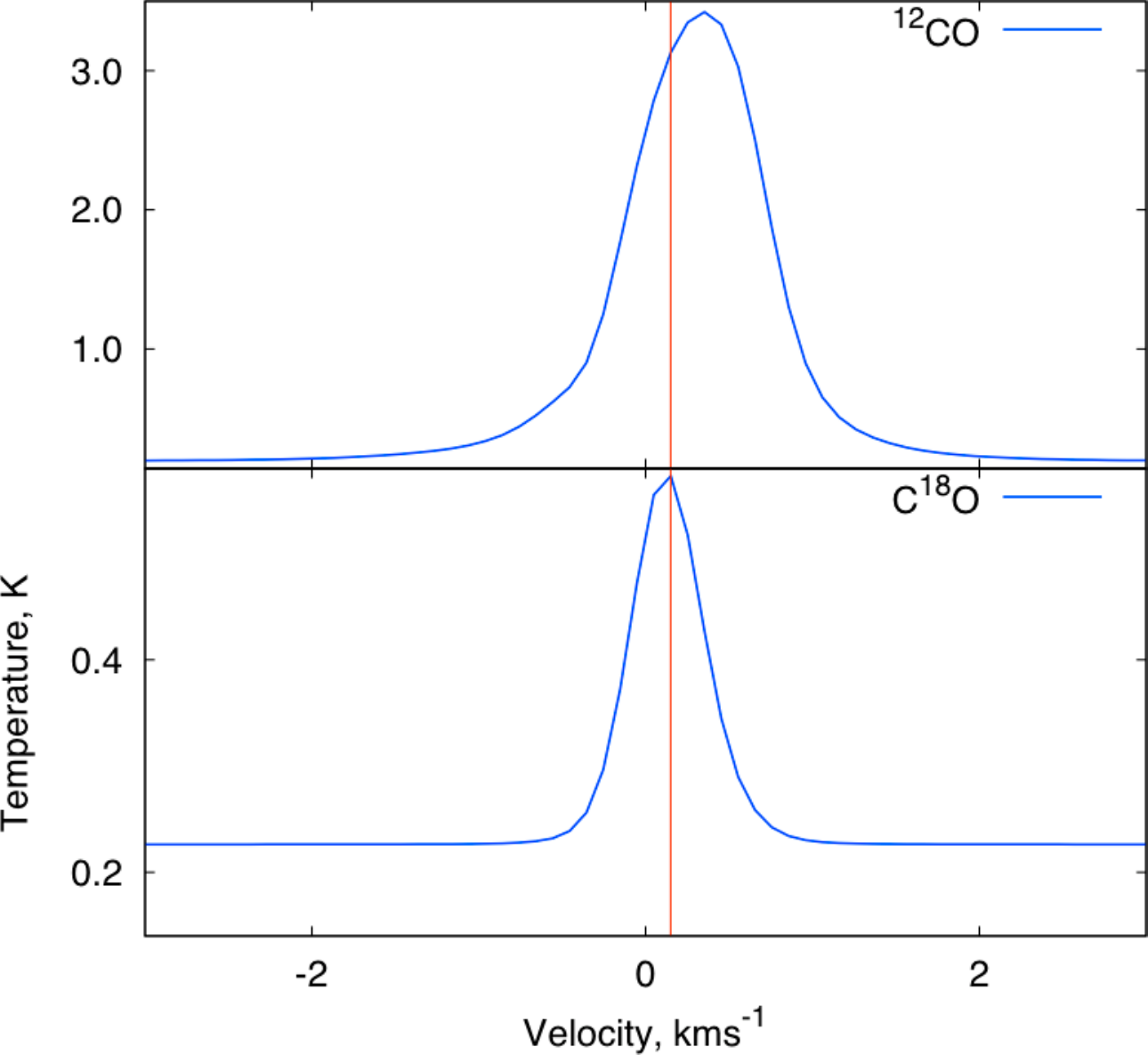}
	\caption{$^{12}$CO and C$^{18}$O line profiles at a viewing angle of $-60$ degrees for the low, medium and high flux models from top to bottom. The vertical lines run through the optically thin line profile peak.}
	\label{inc60}
\end{figure}

In the low flux model (the top panel of Figure \ref{VvsI}) the optically thin C$^{18}$O line {is constant as a function of viewing angle}, whereas the optically thick line varies in velocity dramatically. As discussed in section \ref{kinematics}, this is because {the shell is optically thin to the the C$^{18}$O line and so the line profile peak comes from the interior cloud at all viewing angles.} {The shell is optically thick to the  $^{12}$CO line meaning that the line profile will change with viewing angle as the motion of the shell along the line of sight changes.} Both the shell and central cloud are identified in the optically thick and thin line profiles, but the relative strengths differ.

The high flux model shows similar behaviour to the low flux model, with the optically thick and thin peaks separated, but it is not as extreme. This is because the shell accumulated before driving into the cloud {had lower momentum than in the other models, meaning that the shell rapidly reached pressure equilibrium with the cloud. The shell only continues to propagate into the cloud due to the rocket motion resulting from a weak photo-evaporative outflow and so the velocities are lower}.

In the medium flux model {the shell is sufficiently dense that it is optically thick to both the $^{12}$CO and C$^{18}$O lines. Therefore both line profile peaks come from a region of the BRC with similar kinematic properties and their peak velocities vary with viewing angle in the same way.}

Examples of $^{12}$CO and C$^{18}$O cloud-averaged profiles at a viewing angle of $-60$ degrees are given in Figure \ref{inc60}. These illustrate the points discussed in this section, showing that the low and high flux {models'} optically thin lines stay centred at low velocity whereas the medium flux peak moves to follow the optically thick line. The medium and high flux model optically thick and thin line peaks are only slightly separated at this inclination, whereas the low flux peaks are widely separated by $2.9$\,km\,s$^{-1}$

\subsubsection{The variation of the optically thin line profile FWHM}
The C$^{18}$O FWHM (which is used to calculate the column density, c.f. equation \ref{Ntot}) remains fairly constant at about $0.5$\,km\,s$^{-1}$ in the low and high flux models. {This is because the optically thin line profile peak is determined by the central cloud at all viewing angles.}

\begin{figure}
	\hspace{-15pt}
	\includegraphics[width=8.8cm]{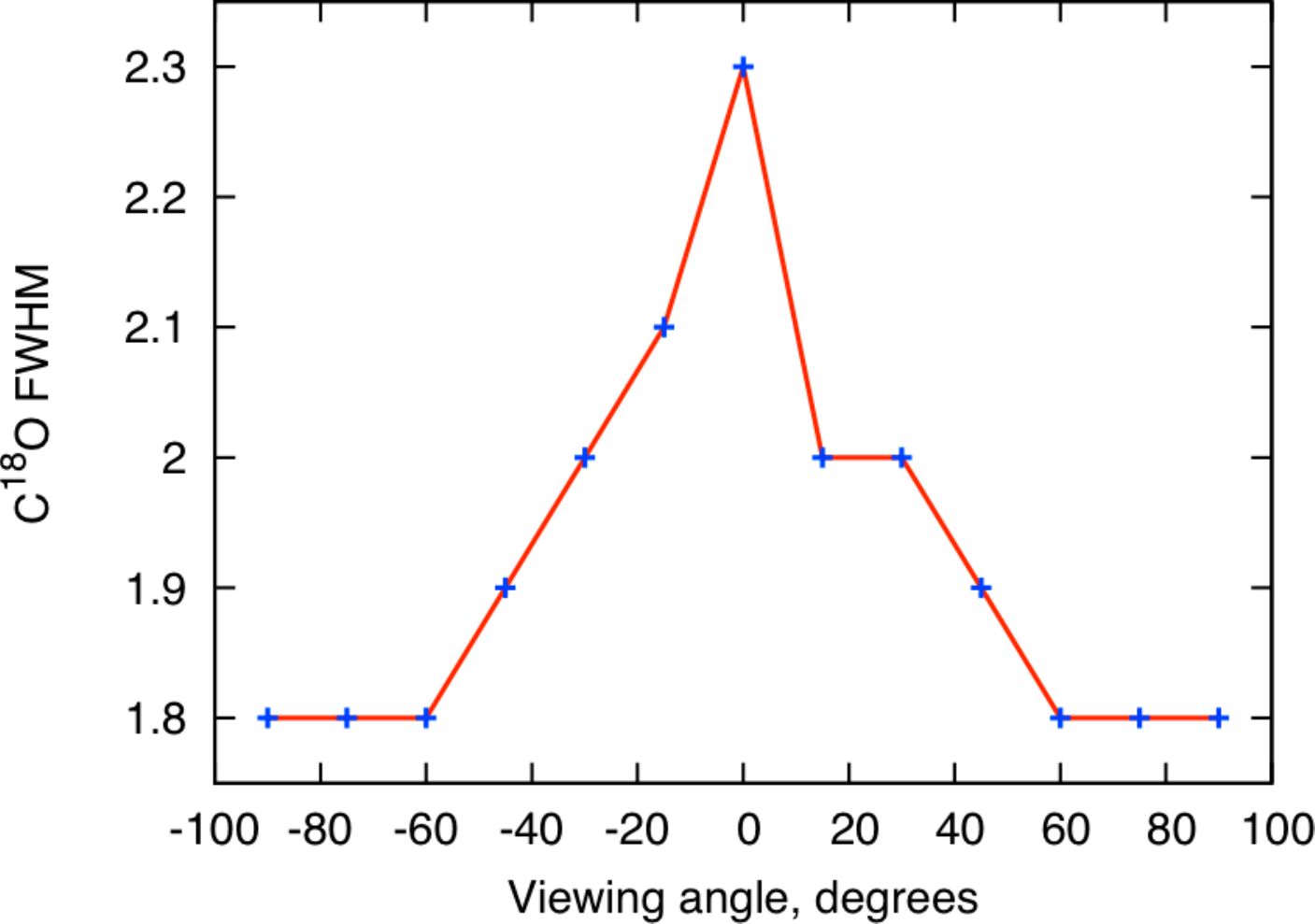}
	\caption{The variation of the medium flux model C$^{18}$O FWHM with viewing angle.}
	\label{FWHMvsI}
\end{figure}

\begin{figure}
	\hspace{-15pt}
	\includegraphics[width=9.2cm]{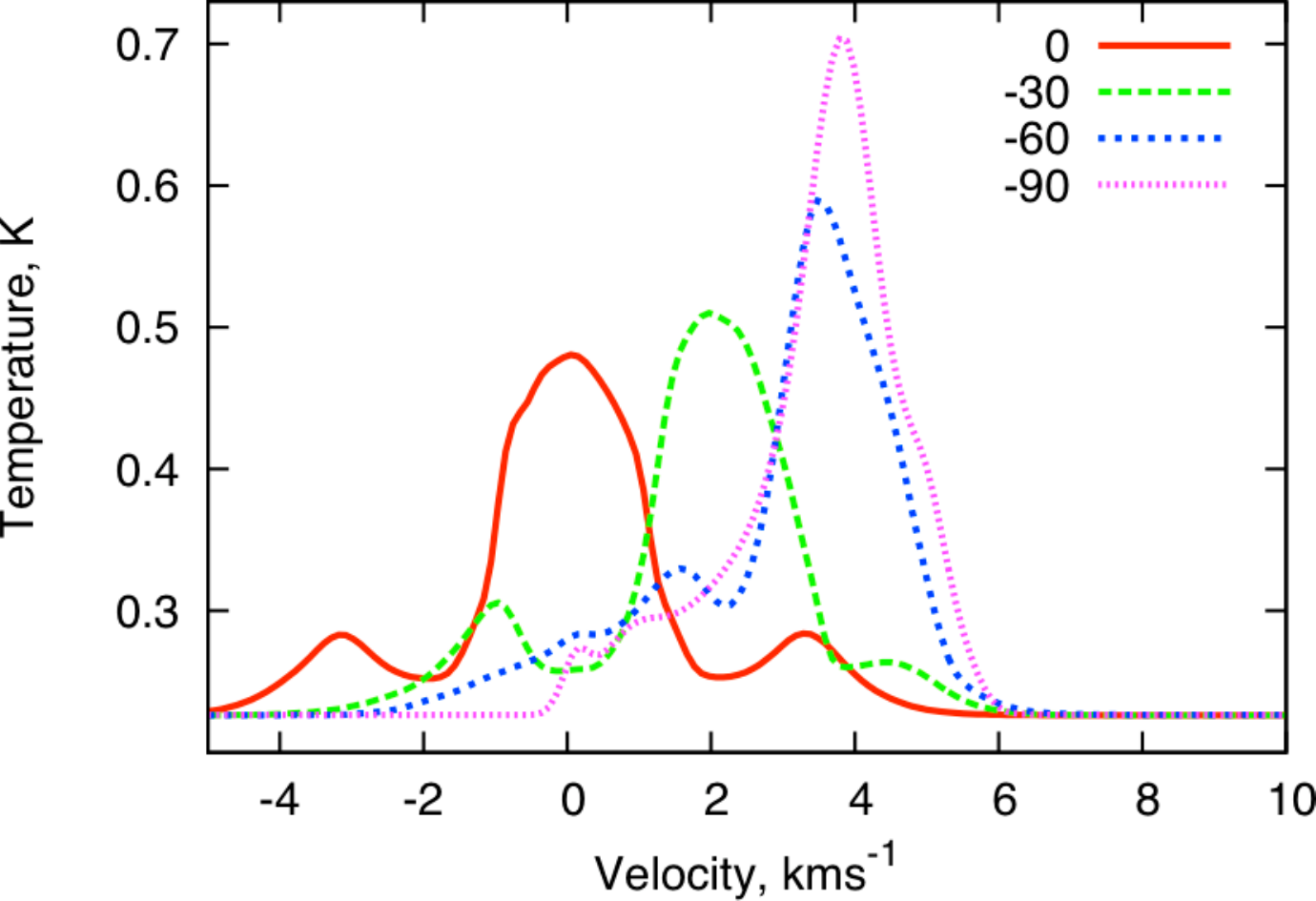}
	\caption{The variation in C$^{18}$O line profile with viewing angle for the medium flux model.}
	\label{incvariation}
\end{figure}

Conversely, the FWHM of the medium flux C$^{18}$O line exhibits a maximum at low viewing angle, decreasing {by up to 20 per cent} as the observer moves to face the object from behind or face on.  This is illustrated in Figure \ref{FWHMvsI}.  The profiles for which these FWHM are calculated are averaged over a constant number of pixels across viewing angles, centred on the area of peak emission and not diluted by the ambient medium. This variation in FWHM is hence not due to varying the size of the region over which the profile is averaged, modifying the size of the line profile peak. Rather, the reason for this variation in the FWHM is {that the shell is optically thick to  C$^{18}$O}. As the observer moves to higher viewing angles a single, {denser component of the shell than that seen edge--on dominates the profile. There is also a smaller distribution of velocities about the peak since only a single shell is contributing to the line profile rather than two.} The result is a {slightly} stronger peak that has a smaller FWHM. This {variation in the line profile} is illustrated in Figure \ref{incvariation}, where the medium flux C$^{18}$O profile is shown as the observer moves from $0$ degrees to $-90$ degrees {in 30 degree intervals}.

\subsubsection{The variation of the line profile symmetry parameter}
\begin{figure}
	\hspace{-9pt}
	\includegraphics[width=8.6cm]{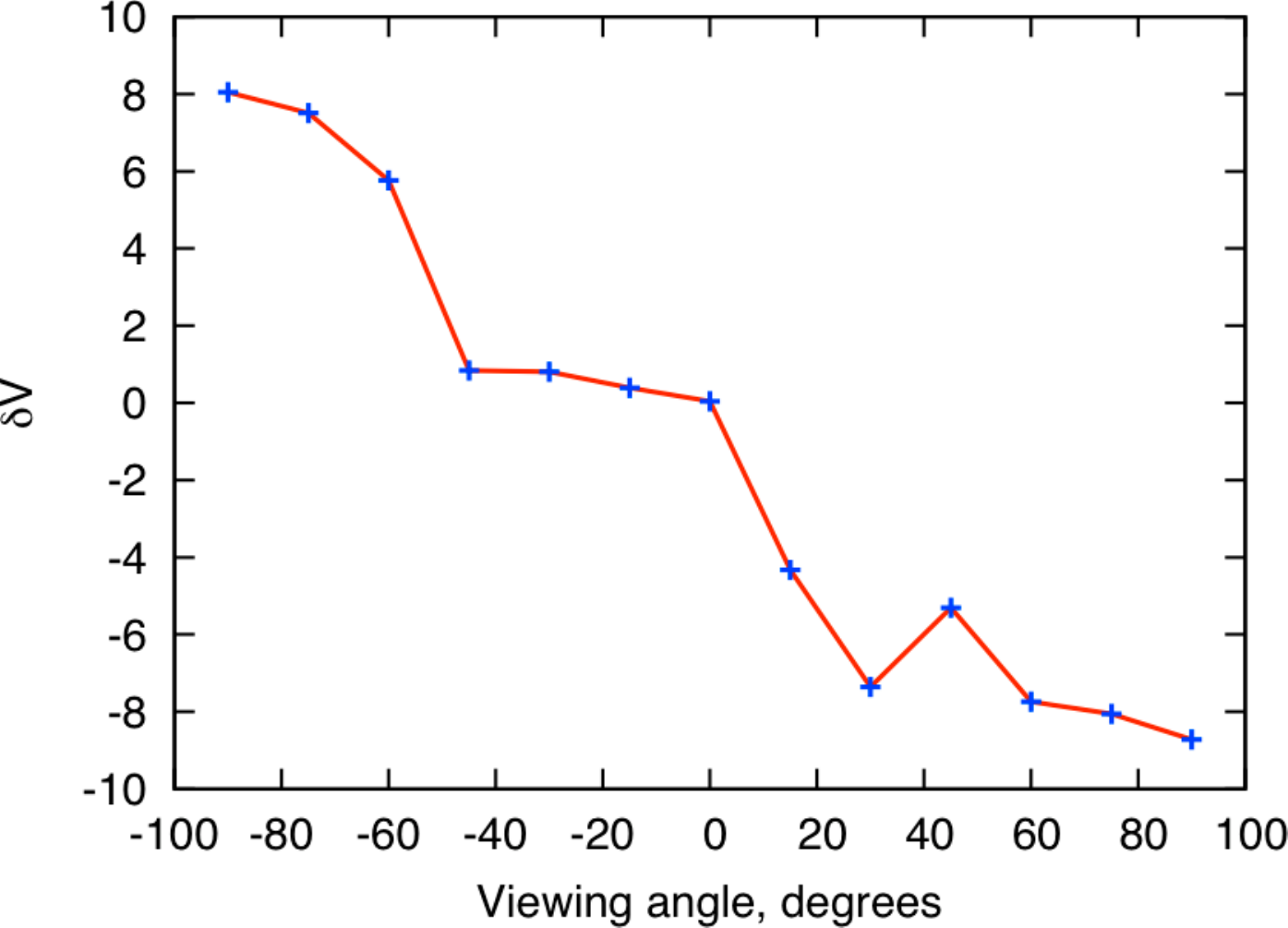}

	\hspace{-14pt}
	\includegraphics[width=8.8cm]{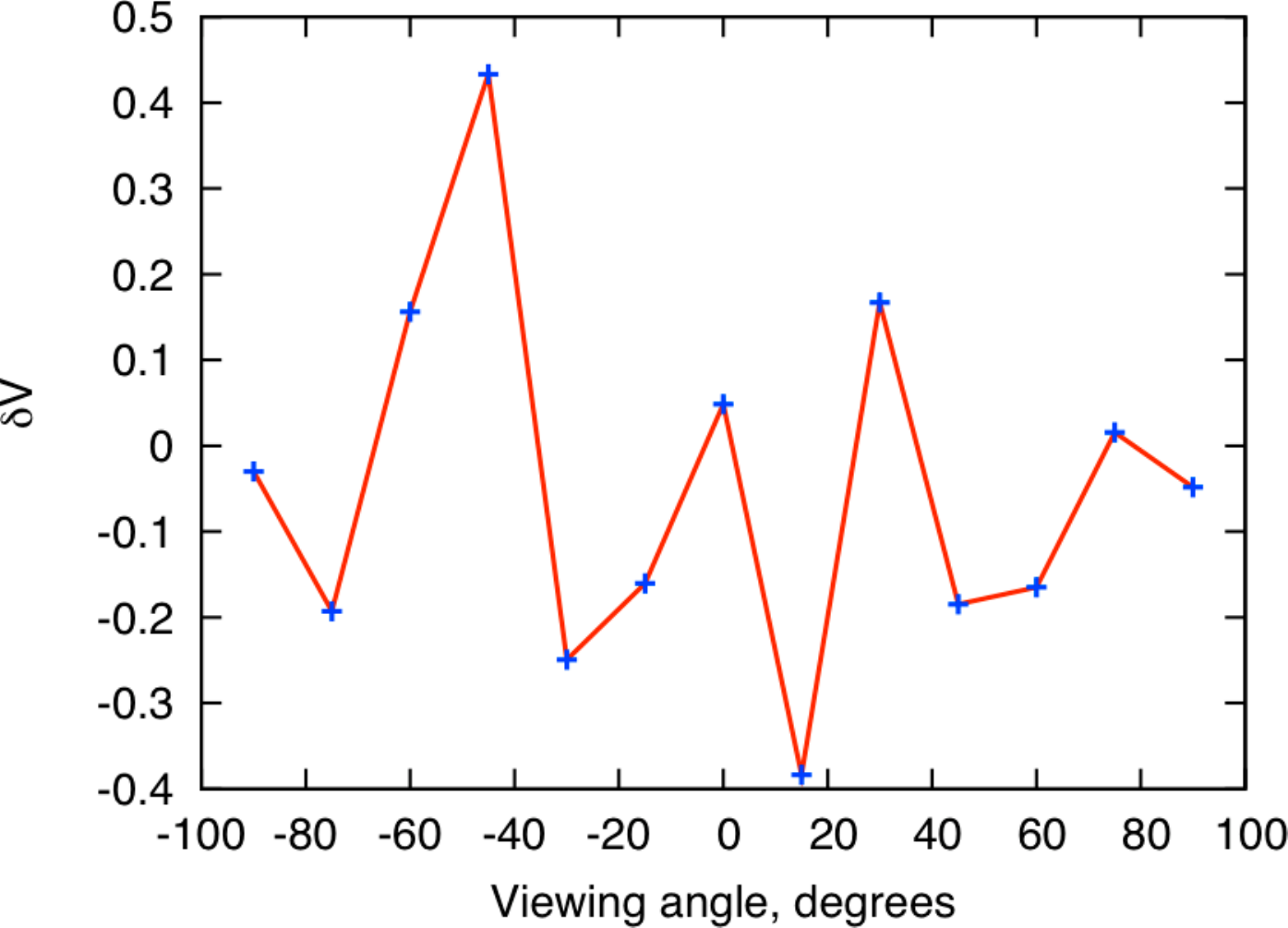}

	\hspace{-19pt}
	\includegraphics[width=9cm]{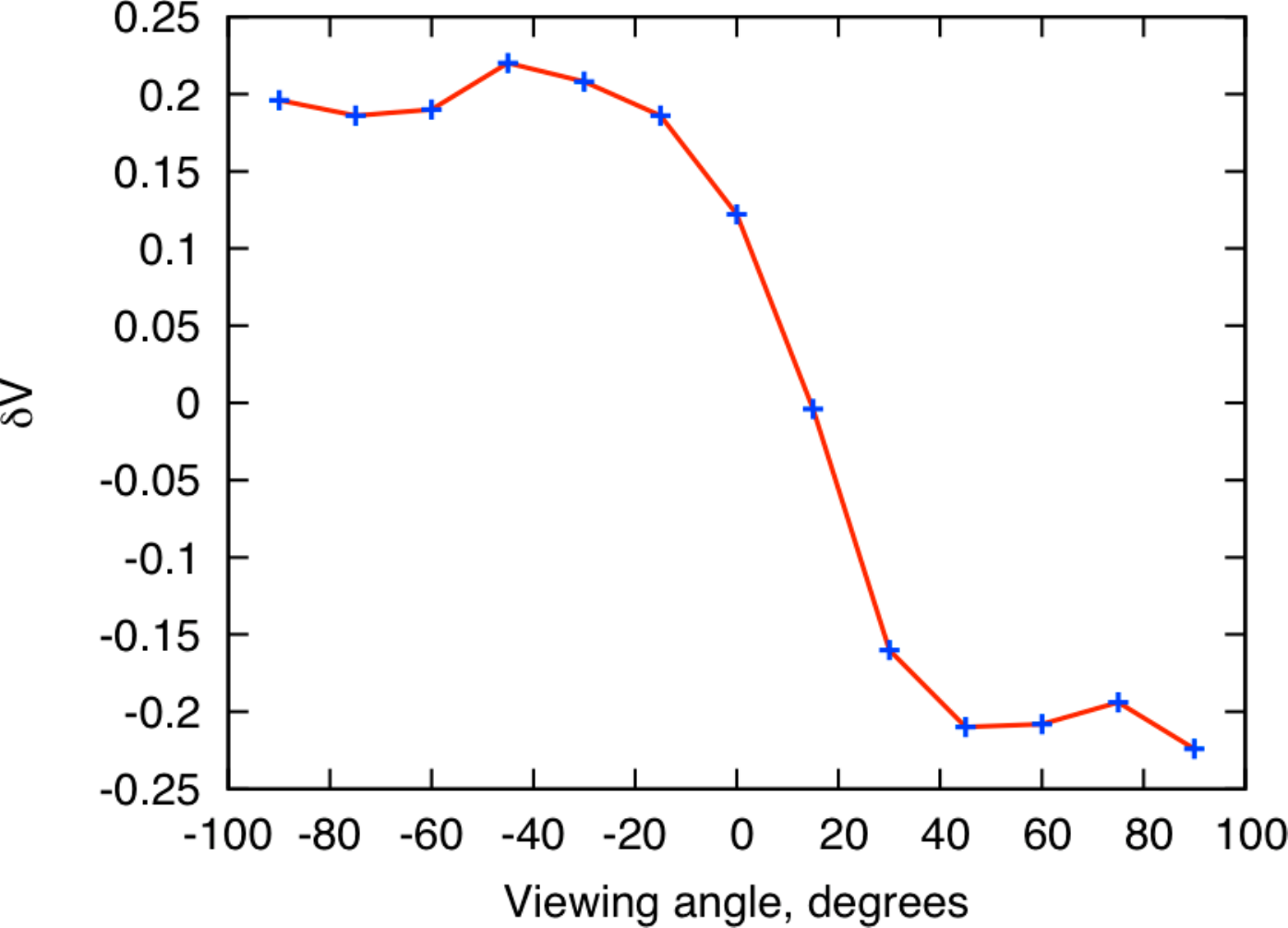}
	\caption{The variation of the profile symmetry with viewing angle for the low (top), medium (middle) and high (bottom) flux models. {The asymmetry parameter $\delta{V}$ (equation \ref{normveldif})} is determined by the difference between the lines of Figure \ref{VvsI} divided by the FWHM of the C$^{18}$O line.}
	\label{dVvsI}
\end{figure}

The symmetric nature of a profile can be quantitatively expressed using the profile symmetry {parameter} $\delta V$ based on an optically thick and an optically thin line, defined as 
\begin{equation}
	\delta{V} = \frac{V_{\rm{thick}} - V_{\rm{thin}}}{\Delta V_{\rm{thin}}}
	\label{normveldif}
\end{equation}
where $V_{\rm{thick}}$, $V_{\rm{thin}}$ and $\Delta V_{\rm{thin}}$ are the source-averaged spectrum peak velocities of the {thick and thin} lines and the FWHM of the optically thin line respectively \citep{1997ApJ...489..719M}. This value gives an indication of the asymmetry in a line profile, with negative values blue-asymmetric and positive values red-asymmetric. \cite{1997ApJ...489..719M} suggest that values in the range $-0.25<\delta{V}<0.25$ should be considered symmetric. {This is usually} applied to a single optically thick line with self absorption to determine whether the red or blue motions are predominantly self absorbed. The optically thin line would have a similar linewidth as the thick line and typically $|\delta$V$| < 1$. In this paper, although both the optically thin and thick lines come from the BRC, they sometimes probe different regions (the shell and the cloud behind the shell) and so the optically thin and thick line peaks may be located at different positions giving rise to larger values of $\delta{V}$. Equation \ref{normveldif} is therefore more of a peak separation function than a symmetry function in this paper, though we still refer to profiles as symmetric or asymmetric depending on the value of $\delta{V}$.
We calculate $\delta$V for all three clouds over each viewing angle to see if there is a systematic variation. The results of this {analysis} are given in Figure \ref{dVvsI}. 

The low flux and high flux models show a systematic transition in $\delta{V}$ with viewing angle. For viewing angles where the observer is looking {face on to the BRC ($<0$ degrees using the convention in Figure \ref{schem})}, $\delta{V}$ shows that there is no or strong red-asymmetry. When the observer is behind the BRC {($>0$ degrees using the convention in Figure \ref{schem})} $\delta{V}$ shows that there is no or strong blue-asymmetry. As already discussed, this is due to the near shell motion (probed by the optically thick line) relative to the interior cloud motion (probed by the optically thin line). The low values of $\delta{V}$ are similar to those found in, for example, \cite{2002ApJ...577..798D}. The large values arise when the shell peak becomes stronger than the {low velocity central cloud peak}. Calculations of $\delta{V}$ have not {yet} been performed for {BRCs} where the optically thick and thin lines are widely separated {and would therefore give rise to larger values}. Some of the profiles given in \cite{2009A&A...497..789U} visually suggest {a large separation} between the optically thick and thin line peaks, such as {those in} SFO 71 and 73.

In the medium flux model there is no clear transition between the dominance of red or blue-asymmetry. This is because the shift in velocity demonstrated in Figure \ref{VvsI} is similar for the optically thin and thick lines {since} they both trace the dense shell region of the cloud. Red and blue-asymmetric profiles can still arise in the medium flux model {(for example at 15 and $-45$ degrees in Figure \ref{dVvsI})}, though typically the profiles are symmetric. The values of $\delta{V}$ obtained for the medium flux cloud are very similar to those obtained for type B-C BRCs in \cite{2002ApJ...577..798D}. 

Given the above, the only way in which a BRC will have a blue-asymmetric profile is if it is viewed from the rear and the shell is not sufficiently dense to dominate the profiles of both the optically thick and thin lines. Requiring that the BRC be viewed from behind {(with a viewing angle $> 0$ degrees using the convention given in Figure \ref{schem}) would reduce} the chances of observing a blue-asymmetric line profile by 50 per cent. This is further reduced depending on the shell densities of real BRCs. These results do therefore provide an explanation for the lack, but not complete absence of, blue-asymmetric BRCs and rather a dominance of symmetric profiles.

\subsection{Molecular cloud conditions}
\label{molcond}
We {applied} the diagnostics detailed in section \ref{calcCond} to each of the clouds to calculate the {mass, temperature and column density at an edge--on inclination (0 degrees using the convention given in Figure \ref{schem})}.  The brightness temperature of the cloud that is used in equation \ref{tauFromT} is the peak of the source-averaged spectrum with the background signal subtracted. We used \textsc{gaia} to obtain a single averaged spectrum for each BRC and fit it with a Gaussian profile to obtain the peak value and FWHM. The radii of {the clouds in} the low, medium and high flux models {were estimated to be} 0.51, 0.19 and 0.8\,pc respectively {from their spatial extent in the simulated images}. 

The inferred properties are all presented in Table \ref{cloud_conditions}, along with the conditions from the model grid for comparison. 
The mass calculated assuming spherical symmetry (equation \ref{M_sph}) is given by M$_{\rm{sph}}$ and the mass calculated by integrating the column density (equation \ref{M_int}) is given by M$_{\rm{int}}$. 

\begin{table*}
\caption{The cloud conditions calculated using $^{12}$CO, $^{13}$CO and C$^{18}$O. {The columns are, from left to right: which model the conditions have been calculated for, the optical depth of $^{13}$CO and C$^{18}$O, excitation temperatures for $^{12}$CO and C$^{18}$O, the column density calculated from synthetic observations, the column density from the model grid, the C$^{18}$O FWHM, the cloud mass calculated assuming spherical symmetry, the cloud mass calculated by integrating the column density over the cloud and the known cloud mass from the model grid. }}
\label{cloud_conditions}
\begin{tabular}{c c c c c c c c c c c}
\hline
Model &  $\tau_{13}$ &  $\tau_{18}$  & $T_{12}$ & $T_{18}$ & $\log(N(H_2))$ & grid $\log(N(H_2))$ & FWHM, $\Delta\nu_{18}$  & M$_{\rm{sph}}$ & M$_{\rm{int}}$ & M$_{\rm{grid}}$\\
 & & & (K) & (K) & (cm$^{-2}$) & (cm$^{-2}$) & (km\,s$^{-1}$) & ($M_\odot$) & ($M_\odot$)  & ($M_\odot$)  \\
\hline
Low        & 2.81 & 0.18 & 12.6 & 6.4 & 21.11 & 21.16 & 0.47  & 100  & 19 & 19 \\
Medium  & 2.56  & 0.16 & 6.4 & 4.5 & 21.61 & 21.37 & 1.88 & 45  & 9 & 4 \\
High       & 1.72  & 0.11 & 6.6 & 6.9 & 20.99 & 20.82 & 0.60 & 185 & 36 & 21\\
\hline
\end{tabular}
\end{table*}

The optical depths are similar to those found observationally by, for example, \cite{2006A&A...450..625U}, \cite{2009MNRAS.400.1726M} and given in \cite{2009A&A...497..789U}.
Other than for $^{12}$CO in the low flux model, the excitation temperatures are consistent underestimates of the prescribed neutral gas temperature of 10\,K. It is often assumed that the BRC is in LTE and that therefore the $^{12}$CO excitation temperature can be used to describe the kinetic temperature of the cloud. The results here suggest that this could be inaccurate by up to a factor of 1.6. The column densities are also similar to those found observationally, for example \cite{2009MNRAS.400.1726M} and \cite{2009A&A...497..789U}. Our values are slightly lower than those from \cite{2006A&A...450..625U}, due to the higher $^{12}$CO excitation (and hence kinetic) temperatures that they obtain, of order 30\,K. \cite{2006A&A...450..625U} attribute this to some internal heating mechanism such as a young stellar object (YSO) or ultra compact (UC) H\,\textsc{ii} region which are not present in our models. Rather, for fully neutral gas we prescribe a {minimum} temperature {in the photoionization calculation} of 10\,K.
The inferred column densities correspond reasonably well to the column density from the model grid, agreeing to within 11, 43 and 32 per cent for the low, medium and high flux models respectively. 

When assuming spherical symmetry, all of the inferred masses are larger than the actual mass in the region over which the diagnostics were performed. The discrepancy ranges from a factor of 5 to 11 under this assumption. For the integrated column density method the agreement is much better, with agreement to within one solar mass in the low flux case up to a factor 2.25 in the {medium} flux case.

{In \cite{2012MNRAS.426..203H} we calculated the mass and temperature of the same clouds using greybody fitting of the cloud SED.} Comparing to the results {here to those from} from \cite{2012MNRAS.426..203H}, calculating the cloud temperature based on greybody fitting of the system SED is a more accurate technique than the molecular line {diagnostics}, typically agreeing to within 1-2\,K. This is because the {former} diagnostic is based on more information, over a larger frequency range from the cloud and also makes fewer assumptions in converting observational intensities to a temperature. For the SED fitting temperature diagnostic, the main assumptions are an index which describes the frequency dependency of dust emissivity and that the SED can be fitted as a greybody. This diagnostic does also probe the whole cloud. However in the molecular line temperature diagnostic (see section \ref{calcCond}, equations \ref{tauFromT} through \ref{c18oExciT}) assumptions include that the optical depths of $^{13}$CO and C$^{18}$O can be related by their abundances (equation \ref{oDepths}), that the cloud is in LTE and that a single temperature applies to the whole cloud. A single line will also only give a diagnostic temperature for the subset of the cloud that it probes.
{The masses calculated} using molecular line diagnostics are more accurate, differing at most by a factor of 2.25 in the integrated column density method compared with a difference of up to a factor of 4 via SED fitting. This is because SED fitting assumes a constant dust to total mass conversion factor between different BRCs. 

A measure of stability against collapse of a BRC is given using the virial theorem, comparing the IBL and neutral cloud pressures \citep{2009apsf.book.....H, 2012MNRAS.426..203H}. In general, the cloud masses have been overestimated here. Given this, the neutral cloud pressure and hence the stability against collapse may also be overestimated when using neutral cloud properties based on molecular line calculations.

\section{Summary and conclusions}
We have generated synthetic molecular line observations of the models of RDI from \cite{2012MNRAS.420..562H}. Using data of the $^{12}$CO, $^{13}$CO and C$^{18}$O (J\,=\,$2\,\rightarrow\,1$) transitions we have analyzed line profiles over the imaged BRCs and replicated standard diagnostics to calculate the BRC properties. Using the derived conditions and line profiles we have searched for signatures of RDI and tested the accuracy of the diagnostics. We have also investigated the variation of BRC line profiles with observer viewing angle. We draw the following main conclusions from this work:
\\

1. The synthetically imaged BRCs have a similar morphology to real BRCs. The optically thin and thick line {integrated intensities} all trace a similar extent of the cloud {in each model}. 
\\

2. The lack of blue-asymmetry observed in BRC line profiles can be explained by the shell of material that drives into the cloud. If the shell is very dense {then it may be optically thick to both $^{12}$CO and C$^{18}$O. If this is the case then the profiles of both lines are dominated by} {emission from the shell and} {have very similar peak velocities that result in a symmetric profile}. In the intermediate case when the shell is less dense the optically thick line profile is dominated by the {high velocity} shell and the optically thin line dominated by the {low velocity} cloud interior to the shell.  {This results in an asymmetric line profile}. {For asymmetric profiles,} when the observer is facing the BRC the shell is moving away into the cloud and there will be a red asymmetry.  If the observer views the BRC from behind then the motion of the shell will be towards the observer and there will be a blue asymmetry. If the shell is sufficiently weak then it will not contribute to the profile and is likely that RDI will not be occurring. 
\\

3. By examining the {known motion of material} in the neutral gas from the model grid we rule out {envelope expansion with core collapse (EECC)} as the cause of the asymmetry in {the simulated line profiles}. This is because expansion from the outer layers of the BRC towards the observer {(a key feature of the EECC model)} is from gas that is ionized, {meaning no molecular gas exists in these regions and they cannot contribute to the line profile}.
\\

4. The profiles that we obtain exhibit shoulders and wings that resemble observations {(see Figure \ref{profiles})}. At edge-on {viewing angles} both {the} near and far shell, as well as the gas interior to the shells, contributes to the profile. This gives rise to more complex profiles with up to three peaks. That such complex profiles exist in observations to date, for example the profiles of SFO 59, 60, 73, 80, 81, 86 and 87 from \cite{2009A&A...497..789U}, is evidence of a shell contributing to the line profile. These systems should be investigated more closely using spatially resolved profiles. At other inclinations the profile is typically either invariant (for the optically thin line) or becomes dominated by a single peak due to the shell with non-Gaussian wings (for the optically thick line). Such profiles are most common in observations due to the higher probability of viewing a BRC at an inclination that is not edge on.
\\


5. For BRCs, failing to identify a PDR {at shorter wavelengths} does not necessarily rule out RDI. If the cloud line profile has a secondary strong blue peak then the shell may be driving towards the observer, something that (according to the models here) only happens if the observer is behind the BRC. As such the PDR is on the opposite side of the cloud so may be more difficult to detect {if the foreground cloud is optically thick}. Examples of this could be {SFO 80, 81 and 86} which were identified as not likely being sites of triggering in  \cite{2009A&A...497..789U} at $8\,\mu\rm{m}$ {(primarily using Midcourse Space Experiment data)}, but have secondary blue peaks. Conversely SFO 87 (with a secondary red peak that suggests the shell is moving away from the observer and the BRC is viewed face on) does have a PDR identified. Analysis of the clouds where no PDR was detected at $8\,\mu\rm{m}$ using longer wavelength data {such as that taken with Herschel or Spitzer} may help to identify a PDR.
\\

6. The cloud conditions that we infer by replicating the diagnostics of, for example \cite{2006A&A...450..625U} and \cite{2009MNRAS.400.1726M}, yield results that are similar to those found observationally. The inferred kinetic temperature differs from the prescribed temperature by up to a factor of 1.6. The column densities for low, medium and high flux models agree with those from the model grid to within 11, 43 and 32 per cent respectively. The cloud masses calculated assuming spherical symmetry are overestimates by up to a factor of 11. Integrating the column density over a region to determine the mass yields much more accurate results, at worst differing from the grid mass by a factor of 2.25 and agreeing more closely for the other models. By comparing with the results from \cite{2012MNRAS.426..203H} we conclude that calculation of cloud temperatures via greybody fitting of the SED is more accurate. However, the mass calculation is more accurate using molecular line diagnostics because the SED fitting assumes a constant dust to total mass conversion factor between clouds. 
\\

This paper marks the third in a series \citep{2012MNRAS.420..562H,2012MNRAS.426..203H} that have attempted to resolve some of the disparity between models and observations of RDI by testing assumptions in the models and simulating observations. At present there are still a number of untested approximations in radiation hydrodynamic modelling, for example the treatment of metals and photodissociation. Star forming regions also have much more complex morphology than the simple case of isolated RDI, which is now being modelled \citep[e.g.][]{2012MNRAS.422.1352D,2012MNRAS.427..625W}. Future simulated observations of these more complex geometries may reveal and resolve further difficulties in translating between theory and observation. {We next intend to build on the work in this paper by using far-infrared observations of BRCs to try and identify triggering in the clouds that we predict are being viewed from behind.}

\section*{Acknowledgments}
The calculations presented here were performed using the University of
Exeter Supercomputer, part of the DiRAC Facility jointly funded by
STFC, the Large Facilities Capital Fund of BIS, and the University of
Exeter. T. J. Haworth is funded by an STFC studentship.  We thank Chris Brunt, Emily Drabek and Jennifer Hatchell for 
useful discussions. {We also thank the referee for their useful comments, which helped to improve the paper.}

\bibliographystyle{mn2e}
\bibliography{molecular.bib}

\appendix

\bsp

\label{lastpage}

\end{document}